\documentclass[twocolumn,lineno]{aastex631}
\usepackage{float} 
\usepackage{amsmath}
\usepackage{placeins}
\usepackage{lipsum}
\usepackage{moreverb}
\usepackage[utf8]{inputenc}
\usepackage{textcomp}
\usepackage{newunicodechar}
\newunicodechar{−}{\textminus}

\newcounter{dummycounter}
\newcommand{\Ha}{\ifmmode \mathrm{H}\alpha \else H$\alpha$\fi}
\graphicspath{{./}{figures/}}

\accepted{June 29, 2026}

\shorttitle{SCALE and a pilot study of Abell 4038}
\shortauthors{Mendes de Oliveira et al.}

\definecolor{myblue}{rgb}{0.12, 0.24, 0.99}
\definecolor{myred}{rgb}{0.99, 0.12, 0.24}

\begin{document}

\title{S-PLUS Clusters And Large-scale Environments (SCALE): \\
I. A catalog of known clusters and groups in DR5 and a pilot study of Abell 4038}

\correspondingauthor{C. Mendes de Oliveira, N. M. Cardoso}
\email{claudia.oliveira@iag.usp.br, natanael.mc@usp.br}

\author[0000-0002-5267-9065]{C. Mendes de Oliveira}
\affiliation{Departamento de Astronomia, Instituto de Astronomia, Geofísica e Ciências Atmosféricas, U. de São Paulo, SP, 05508-090, Brazil}

\author[0000-0002-2238-9665]{N. M. Cardoso}
\affiliation{Escola Politécnica, Universidade de São Paulo, São Paulo, SP, 05508-010, Brazil}

\author[0000-0003-2540-7424]{P. A. A. Lopes}
\affiliation{Observatório do Valongo, Universidade Federal do Rio de Janeiro, Ladeira do Pedro Antônio 43, Rio de Janeiro, RJ 20080-090, Brazil}

\author[0000-0002-3813-2565]{A. L. B. Ribeiro}
\affiliation{Laboratório de Astrofísica Teórica e Observacional, Universidade Estadual de Santa Cruz, Ilhéus, BA 45662-900, Brazil}

\author[0000-0001-9243-3425]{D. E. Olave-Rojas}
\affiliation{Departamento de Tecnolog\'ias Industriales, Facultad de Ingenier\'{i}a, Universidad de Talca, Los Niches km 1, Curic\'o, Chile}

\author[0000-0003-4630-1311]{A. Krabbe}
\affiliation{Departamento de Astronomia, Instituto de Astronomia, Geofísica e Ciências Atmosféricas, U. de São Paulo, SP, 05508-090, Brazil}

\author[0000-0002-3876-268X]{L.  Sodré Jr.}
\affiliation{Departamento de Astronomia, Instituto de Astronomia, Geofísica e Ciências Atmosféricas, U. de São Paulo, SP, 05508-090, Brazil}

\author[0000-0003-3921-2177]{R. Demarco}
\affiliation{Institute of Astrophysics, Facultad de Ciencias Exactas, Universidad Andr\'es Bello, Sede Concepci\'on, Talcahuano, Chile}

\author[0009-0007-2396-0003]{A. V. Smith Castelli}
\affiliation{Instituto de Astrofísica de La Plata, CONICET-UNLP, Paseo del Bosque s/n, B1900FWA, La Plata, Argentina}
\affiliation{Facultad de Ciencias Astronómicas y Geofísicas, U. Nacional de La Plata, Paseo del Bosque s/n, B1900FWA, La Plata, Argentina}

\author[0000-0001-9672-0296]{R. Cid Fernandes}
\affiliation{Departamento de F\'isica, Universidade Federal de Santa Catarina, Florian\'opolis, SC  88040-900, Brazil}

\author[0000-0001-7907-7884]{F. R. Herpich}
\affiliation{Laboratório Nacional de Astrofísica (LNA/MCTI), Rua Estados Unidos, 154, Itajubá, MG 37504-364, Brazil}

\author[0000-0002-7005-8983]{S. Torres-Flores}
\affiliation{Departamento de Astronomía, Universidad de La Serena, Avda. Ra\'ul Bitr\'an 1305, La Serena, Chile}

\author[0000-0002-7272-9234]{E. R. Carrasco}
\affiliation{International Gemini Observatory, NFS NOIRLab, Casilla 603, La Serena, Chile}

\author[0000-0002-6268-8600]{E. V. R. Lima}
\affiliation{Departamento de Astronomia, Instituto de Astronomia, Geofísica e Ciências Atmosféricas, U. de São Paulo, SP, 05508-090, Brazil}

\author[0009-0003-6609-1582]{G. Oliveira Schwarz}
\affiliation{Escola Politécnica, Universidade de São Paulo, São Paulo,  SP, 05508-010, Brazil}

\author[0000-0001-9523-5272]{A. P. Costa}
\affiliation{Laboratório de Astrofísica Teórica e Observacional, Universidade Estadual de Santa Cruz, Ilhéus, BA 45662-900, Brazil}

\author[0000-0001-8450-5193]{L. Doubrawa}
\affiliation{Department of Physics, University of Helsinki, P.O. Box 64, FI-00014 Helsinki, Finland}
\affiliation{Departamento de Astronomia, Instituto de Astronomia, Geofísica e Ciências Atmosféricas, U. de São Paulo, SP, 05508-090, Brazil}

\author[0009-0003-1364-3590]{G. P. Montaguth}
\affiliation{Departamento de Astronomia, Instituto de Astronomia, Geofísica e Ciências Atmosféricas, U. de São Paulo, SP, 05508-090, Brazil}

\author[0009-0006-0373-8168]{C. Lima-Dias}
\affiliation{Departamento de Astronomía, Universidad de La Serena, Avda. Ra\'ul Bitr\'an 1305, La Serena, Chile}

\author[0000-0001-9196-7038]{E. S. Cypriano}
\affiliation{Departamento de Astronomia, Instituto de Astronomia, Geofísica e Ciências Atmosféricas, U. de São Paulo, SP, 05508-090, Brazil}

\author[0009-0008-9042-4478]{M. S. Carvalho}
\affiliation{Departamento de Astronomia, Instituto de Astronomia, Geofísica e Ciências Atmosféricas, U. de São Paulo, SP, 05508-090, Brazil}

\author[0000-0003-2415-3338]{C. Lobo}
\affiliation{Instituto de Astrof\'isica e Ci\^encias do
Espa\c{c}o, CAUP, Universidade do Porto, Rua das Estrelas, 4150-762 Porto,
Portugal}
\affiliation{Departamento de F\'{\i}sica e Astronomia, Faculdade de Ci\^encias,
Universidade do Porto, R. Campo Alegre 687, 4169-007 Porto, Portugal}

\author[0000-0002-7865-3971]{M. Fonseca-Faria}
\affiliation{Laboratório Nacional de Astrofísica (LNA/MCTI), Rua Estados Unidos, 154, Itajubá, MG 37504-364, Brazil}

\author[0000-0001-6480-1155]{L. Nakazono}
\affiliation{Observatório Nacional, Rua Gal. José Cristino 77, 20921-400, Rio de Janeiro, RJ, Brazil}

\author[0000-0002-6164-5051]{A. R. Lopes}
\affiliation{Departamento de Astronomia, Instituto de Astronomia, Geofísica e Ciências Atmosféricas, U. de São Paulo, SP, 05508-090, Brazil}
\affiliation{Instituto de Astrofísica de La Plata, CONICET-UNLP, Paseo del Bosque s/n, B1900FWA, La Plata, Argentina}

\author[0000-0002-8048-8717]{F. Almeida-Fernandes}
\affiliation{Instituto Nacional de Pesquisas Espaciais, \\ Av. dos Astronautas 1758, Jardim da Granja,12227-010 S\~ao Jos\'e dos Campos, SP, Brazil}

\author[0000-0002-2484-7551]{A. Kanaan}
\affiliation{Departamento de F\'isica, Universidade Federal de Santa Catarina, Florian\'opolis, SC  88040-900, Brazil}

\author[0000-0002-0138-1365]{T. Ribeiro}
\affiliation{Rubin Observatory Project Office, 950 N. Cherry Ave, Tucson 85719, USA}

\author[0000-0002-4064-7234]{W. Schoenell}
\affiliation{The Observatories of the Carnegie Institution for Science, 813 Santa Barbara St, Pasadena, CA 91101, USA}





\begin{abstract}


Within the framework of the Southern Photometric Local Universe Survey (S-PLUS), we introduce {\bf S}-PLUS {\bf C}lusters {\bf A}nd {\bf L}arge-scale {\bf E}nvironments (SCALE), a project dedicated to the study of galaxy clusters, groups, and their environments using 12-band photometry of S-PLUS combined with spectroscopic and photometric data from the literature. In this first paper, we present a catalog of 83 previously known systems in the redshift range $0.008 \leq z_{\rm spec} \leq 0.1$, for which we derive $R_{200}$, $M_{200}$, and velocity dispersions. Spectroscopic members are selected and matched with S-PLUS photometric redshifts (photo-$z$s). We find very good agreement between literature spectroscopic redshifts (spec-$z$s) and S-PLUS photometric redshifts (photo-$z$s), demonstrating the potential of the latter for cluster and group membership determination. As a proof of concept, we obtain photometric memberships for Abell\,4038 using the Reliable Photometric Membership technique. A two- and three-dimensional analysis of the region within $10 h^{-1}$ Mpc ($10\times R_{200}$) from the center of Abell\,4038 reveals about a dozen substructures including two additional clusters within $1.3\times R_{200}$ (Abell\,4038B and Abell\,4049). A color–luminosity segregation analysis shows that more luminous (less luminous) galaxies are redder (bluer),  as expected. Low-concentration galaxies ($C \leq 2.5$) exhibit a weaker color–luminosity dependence, compared to higher-concentration ones, indicating mass-dependent evolutionary pathways that challenge a simple morphology–color dichotomy, with low-luminosity galaxies presenting bluer colors largely independent of concentration. The SCALE catalog provides a valuable basis for future studies of large-scale structures and their connection to galaxy evolution.

\end{abstract}



\section{Introduction} \label{sec:intro}

Our understanding of how environment drives galaxy evolution has advanced rapidly over the past decades with the advent of wide-field spectroscopic and photometric surveys, e.g. the Sloan Digital Sky Survey \citep[SDSS;][]{York2000}, Six-Degree Field \citep[6dF;][]{6df,6df-dr3}, Pan-STARRS \citep{Kaiser+02}, the Dark Energy Survey \citep[DES;][]{DES+16}, and the 
Dark Energy Spectroscopic Instrument (DESI) Legacy Imaging Survey \citep[Legacy;][]{legacy}. These surveys enabled the construction of samples of voids, galaxy pairs, groups, and clusters spanning a wide range of physical properties, revealing environmental mechanisms operating across different cosmic scales \citep{lewis202,2003Lambas,2004Kauffmann,Rojas_2005,2014Tempel,2024Wen}.

A key limitation of many previous studies is the difficulty in determining reliable memberships, particularly for faint galaxies. Except for Stripe 82 \citep{2013Dawson,2014Annis} and the very nearby systems, spectroscopic samples have relatively bright absolute magnitude limits. For instance, SDSS, with a spectroscopic limit of $r = 17.7$ mag, probes the dwarf regime only in very local structures ($z \sim 0.02$).

This situation has improved with the Southern Photometric Local Universe Survey (S-PLUS) DR5 (Lima et al. 2026, submitted) and the 
DESI DR1 \citep{desi-dr1}, which provide precise photometric and spectroscopic redshifts over large sky areas, extending reliable distance measurements to galaxies with $r \sim 18$–20 mag. Combined with SDSS when necessary, these datasets enable studies of nearby systems out to $5$–$10\times R_{200}$, including low-mass galaxies (down to $M_r = -13$ in the nearest cases). They also allow the construction of new cluster catalogs based on eROSITA detections (Doubrawa et al. 2026, in prep.), including members to $r = 21$ mag and extending beyond the virial radius.

DESI \citep{desi} is a wide-field, highly multiplexed spectrograph on the 4-m Mayall Telescope at Kitt Peak National Observatory. Its five-year survey will obtain tens of millions of redshifts over $\sim$14,000 deg$^2$. The first data release (DR1; \citealp{desi-dr1}) provides high-confidence redshifts for 18.7\,M objects, including galaxies, stars, and quasars. Of these, DR1 includes 1,065,595 galaxies with redshift in range 0.08 to 0.1.

S-PLUS \citep{MendesdeOliveira+19} is a wide-field imaging survey mapping 9,300 deg$^2$ of the southern sky with a dedicated 80-cm telescope at Cerro Tololo Interamerican Observatory (CTIO). It employs a unique 12-band filter system combining broad, medium, and narrow bands centered on key spectral features, enabling detailed sampling of galaxy spectral energy distributions. These data yield accurate photometric redshifts (0.5–3\%;  \citealp{Molino+20,Lima+22}), improving cluster membership determination and stellar population studies. Even in regions covered by spectroscopic surveys such as SDSS and DESI, incompleteness remains significant due to fiber collisions and the difficulty of observing faint passive galaxies. In this context, S-PLUS provides a crucial complementary dataset. In DR5, catalogs covering 4,500 deg$^2$ include photo-$z$s for millions of galaxies down to $r = 21$ mag (Lima et al. 2026, submitted).

The combination of DESI spectroscopy for bright galaxies and S-PLUS photo-$z$s for fainter systems enables a comprehensive reassessment of nearby groups and clusters ($z < 0.1$) out to large clustercentric distances. This strategy is particularly powerful for studying infall regions and environmental mechanisms such as galaxy preprocessing and post-processing.

Here, we introduce {\bf S}-PLUS {\bf C}lusters {\bf A}nd {\bf L}arge-scale {\bf E}nvironments (SCALE), a program aimed at studying galaxy systems using homogeneous S-PLUS imaging combined with available spectroscopy and deep imaging.  Information about the project, including publications, online supplementary material, catalogs, and code, are available at \url{https://splus-scale.github.io}.

In this first paper in the SCALE series, we revisit a sample of 83 well-known groups and clusters with $0.008 \le z \le 0.10$, selected from the literature based on spectroscopic redshifts. For these systems, we compute $M_{200}$, $R_{200}$, velocity dispersion, and determine the spectroscopic membership. As a case study, we analyze the field of \object{Abell 4038} or \object{A4038} (throughout this paper, we will refer to Abell 4038 interchangeably as A4038), performing substructure detection and a luminosity–color–concentration analysis.

The catalog presented in Doubrawa et al. (2026, submitted; Paper II) extends this effort by performing a systematic detection of galaxy clusters in S-PLUS DR5 via positional matches with eROSITA X-ray sources \citep{merloni+24}. Paper I, the present work, establishes a well-characterized reference sample of known systems and Paper II applies similar tools to a broader, X-ray-selected cluster population, providing a complementary view of galaxy systems and their large-scale environments.

An early scientific application within the SCALE framework was presented by \cite{Doubrawa+25}, who used S-PLUS DR5 photometry to investigate the alignment between brightest cluster galaxies (BCGs) and their satellite populations in 58 Stripe 82 \citep{2013Dawson,2014Annis} clusters ($z < 0.12$). They found that the alignment signal is stronger in systems hosting redder and more luminous BCGs, highlighting the role of both cluster-scale dynamics and the surrounding large-scale structure in shaping galaxy orientations. While that study explored galaxy alignments using a subsample of SCALE clusters, the present paper and Paper II establish the homogeneous cluster sample and derived properties that define the broader goals of the program.                                                          

The present paper is organized as follows. Section \ref{sec:Sample} presents the 83 S-PLUS DR5 clusters with spectroscopic redshifts \citep{Lima+24}, the selection of members and interlopers to $5 \times R_{200}$, and the calculation of $M_{200}$, $R_{200}$, and velocity dispersion. Section \ref{sec:ErikANDVitor-Section} evaluates the precision of S-PLUS photo-$z$s and their use in membership determination, illustrated for A4038. Section \ref{sec:Andre-Daniela-Section} presents the substructures in the region around A4038 and an analysis of the luminosity-color-concentration of galaxies in the cluster. Sections \ref{sec:discussion} and \ref{sec:conclusions} discuss the results and summarize the conclusions.

Throughout this paper, we adopt a cosmology with $\Omega_{M}=0.3$, $\Omega_{\Lambda}=0.7$, and $H_{0}=100$ {\it h} km s$^{-1}$ Mpc$^{-1}$, with {\it h}=0.7. Unless otherwise stated, uncertainties are quoted at the $1\sigma$ level.

\section{Sample, properties and spectroscopic member selection} 
\label{sec:Sample}

The sample combines clusters and groups from the NASA's Archive of Data on Energetic Phenomena (HEASARC) catalogs\footnote{\url{https://heasarc.gsfc.nasa.gov}} and the compilation of 1,844 X-ray systems by \citet{Yuan+22}. From these, we selected all clusters within the S-PLUS DR5 Main Survey footprint that have spectroscopic data available in the literature and nominal redshifts in the range $0.008 \le z \le 0.10$. The upper limit ensures a reliable determination of the photometric and stellar population parameters, since most members are well imaged with high signal-to-noise ratio in the Legacy Survey and/or in S-PLUS. This allows robust nonparametric photometric fits and spectral energy distribution analysis across the 12 filters (when S-PLUS data are available), results that will be presented in a companion paper (Mendes de Oliveira et al. in prep.). The lower limit avoids very nearby systems where the photo-$z$ estimates, derived by machine-learning methods, have larger fractional errors \citep{Lima+22} that could bias the results.

Spectroscopic data were obtained from the compilation by \citet{Lima+24}, which collects most published galaxy spectroscopy from multiple sources, in the Southern Hemisphere  (declinations $\le +$5 deg). The used version of the compilation contained data from more than 5,000 catalogs, providing redshift information for 16 million objects among galaxies, stars, and quasars. More details about this compilation can be found on its GitHub page\footnote{\url{https://github.com/ErikVini/specz_compilation}}, and is publicly available on Zenodo\footnote{\url{https://zenodo.org/records/15127060}}. 

In order to draw an initial sample of cluster and group galaxies, objects with spectroscopic redshifts within $\Delta z = \pm 0.01667$ of the system´s cataloged mean redshift\footnote{This corresponds to $\pm$ 5000 km s$^{-1}$, a velocity range that is large enough for membership selection even for the most massive clusters.}, and located within a projected radius of $10~h^{-1}$ Mpc from the cluster center, were selected. After that, membership was determined using the shifting-gapper technique \citep{Fadda+96,Lopes+09}, applied to all galaxies with available spectroscopic redshifts in the vicinity of each system.

The original coordinates of the groups and clusters in the sample come from the list of \citet{Yuan+22}\footnote{\url{http://zmtt.bao.ac.cn/galaxy_clusters/dyXimages/combined.html}} or from the HEASARC archive, but when the  shifting-gapper technique \citep{Fadda+96,Lopes+09} was used for the first time, the galaxy distribution in the projected phase-space (PPS) was inspected and small adjustments in the cluster/group centers were made. To ensure a robust dynamical characterization and maximize the symmetry of the velocity distribution, the central coordinates and redshifts were refined by re-estimating them from member galaxies within $0.30~h^{-1}$ Mpc. In addition, for the initial estimate of the cluster velocity dispersion ($\sigma_{\text{cl}}$), only members within $2.5~h^{-1}$ Mpc were considered. For each cluster, $M_{200}$ was estimated using equation 1 of 
\citet{Ferragamo+20} \citep[but also see][]{Munari+13} 
applied to the selected members. Corrections to $\sigma_{\text{cl}}$ and $M_{200}$ introduced by \citet{Ferragamo+20} were also implemented. From these mass estimates, values for $R_{200}$ were derived, following the procedures detailed in \citet{Lopes+09,Lopes+14,Lopes+18,Lopes+24} and \citet{Ferragamo+20}. Values for $\sigma_{\text{cl}}$ and $M_{200}$ were then recalculated, now restricting the analysis to members within $R_{200}$, instead of $2.5~h^{-1}$ Mpc, to obtain a new estimate of $R_{200}$. This iterative process was repeated until $R_{200}$ varied by less than 10\%. This refinement, compared to the previous version of the code (see \citealt{Lopes+24}), had no significant impact on the final estimates of the cluster/group properties ($\sigma_{\text{cl}}$, $M_{200}$, and $R_{200}$).

The final sample comprises 83 systems with masses ranging from $1.1 \times 10^{12} M_{\odot}$ to $1.75 \times 10^{15} M_{\odot}$ (Table \ref{tab:59clusters}).
Table \ref{tab:59clusters} lists the identification of each system, central coordinates, redshift, $\sigma_{\text{cl}}$, $R_{200}$, $M_{200}$, the number of spectroscopic cluster members within $10~h^{-1}$ Mpc, the number of spectroscopic members within $R_{200}$, and the small offset $s$ between photo-$zs$ and the spectroscopic redshifts for that specific cluster/group (see Section \ref{sec:photo-z-spec-z-comparison}). For systems with few members, and particularly those with $M<10^{13} M_{\odot}$, where virialization is unlikely, the measurements are highly uncertain and will require future more realistic determinations. 

To estimate $M_{200}$ uncertainties consistently across the sample, the statistical error on the line-of-sight velocity dispersion, $\sigma_{\rm cl}$ was propagated by accounting for the individual galaxy radial-velocity uncertainties, and added in quadrature to the intrinsic scatter of the $\sigma$–$M$-scaling relation. 
For each system, $N_{\rm gal}=N_{200}$ was adopted, since $\sigma_{\rm cl}$ and $M_{200}$ were obtained iteratively using only members within the current estimate of $R_{200}$. The statistical uncertainty on $\sigma_{\rm cl}$ was calculated as a function of $N_{\rm gal}$ using the variance calibration of \citet{Ferragamo+20} (applied to the standard deviation, adopted here as the velocity-dispersion estimator) and propagated to mass using the inverted relation $\sigma$--$M$ \citep{Munari+13}. An intrinsic scatter term for galaxy tracers was also included (taken as a 12\% scatter in $\sigma_{\rm cl}$; \citealt{Munari+13}), which sets a floor to the achievable precision. In our sample, the typical mass uncertainty (median of $\Delta M_{200}/M_{200}$) is $\simeq 38\%$ (i.e $\simeq 0.17$ dex), but increases sharply for poorly sampled systems; for systems with $N_{200}<10$, which in our sample correspond predominantly to the lowest-mass halos ($M_{200}\lesssim 10^{13}\,M_{\odot}$), $\Delta M_{200}/M_{200}$ goes from $\sim 70\%$ to $140\%$, exceeding 100\% in two cases (NGC3054 and [YMV2007]7609). This reflects both small-number statistics and non-linear amplification when converting $\sigma_{\rm cl}$ into $M_{200}$. Given these uncertainty levels, dynamical masses below $M_{200}\lesssim 10^{13}\,M_{\odot}$ should be used with particular care. More robust estimates will require additional spectroscopy to increase $N_{200}$ and/or independent mass proxies (e.g. X-ray/SZ, richness, or weak lensing) that are less sensitive to small-number statistics.

Given the substantial uncertainties that affect the dynamical mass estimates, a direct comparison with independent mass proxies provides a valuable consistency check. In the following, the M$_{200}$ estimates are compared with masses derived from eROSITA X-ray observations \citep{Bulbul2024}.
As the eROSITA catalog provides masses defined within $R_{500}$, $M_{500}$ values are converted to $M_{200}$ assuming a Navarro-Frenk-White halo profile \citep{Navarro1996} with a fixed concentration $c=4.5$ \citep{Duffy2008}. 
Matching the cluster and group samples resulted in $49$ systems with $\Delta R<0.5~h^{-1}$ Mpc and $\Delta z < 0.005$. 
The agreement between the two mass estimates is quantified using a logarithmic mass ratio. The distribution of the ratio shows a small median offset of $-0.03$ dex, indicating that dynamical masses are on average $\sim8\%$ lower than the eROSITA X-ray masses. The observed scatter has a $\sigma_{MAD}=0.20$ dex (defined as $median(|X_i-\bar{X}|)$, where $\bar{X}$ is the data median), consistent with expectations from comparisons between dynamical and X-ray mass proxies. By propagating the individual mass uncertainties, an intrinsic scatter of $\sigma_{int}=0.27$~dex is inferred. These values are reduced to $\sigma_{MAD}=0.10$ dex and $\sigma_{int}=0.1$ dex when only structures with at least 50 galaxies within $R_{200}$ are considered. More details are provided in Appendix \ref{sec:mass-comparison}. Overall, the small median offset and moderate intrinsic scatter indicate that the two mass estimates are statistically consistent within their expected uncertainties.

Table 2 (available in the online version only; \url{https://splus-scale.github.io/publications/1}) lists the spectroscopic members and interlopers within a projected radius of $5 R_{200}$ of each of the 83 clusters and groups of Table \ref{tab:59clusters} \citep[with spectroscopic redshifts compiled by][]{Lima+24}, together with more than 50 galaxy parameters derived in this work or collected from the literature. Galaxies classified as members using the shifting-gapper technique are assigned \texttt{flag\_member} = 0, while interlopers (with spectroscopic redshifts within $\Delta z = \pm 0.01667$ of the cluster mean redshift) are assigned \texttt{flag\_member} = 1 in Table 2.

\startlongtable
\begin{deluxetable*}{lrrlcccrrc}

\tablecaption{General properties of the 83 S-PLUS clusters and groups in the sample presented in this work, ordered by increasing spectroscopic redshift, from $z = 0.008$ to $0.1$. The coordinates (J2000) are in degrees. $N_\mathrm{Tot}$ is the total number of cluster members, selected as described in Section \ref{sec:Sample}, including those beyond the cluster virial radius (inside $10\ h^{-1}$ Mpc). $N_{200}$ is the number of cluster members within a radius of $R_{200}$, and $s$ gives the redshift offset described in Section \ref{sec:photo-z-spec-z-comparison}.}

\label{tab:59clusters}

\tablehead{
\colhead{Cluster/Group} & 
\colhead{R.A.} & 
\colhead{Dec.} & 
\colhead{$z$} & 
\colhead{$\sigma_{cl}$} & 
\colhead{$R_{200}$} & 
\colhead{$M_{200}$} & 
\colhead{$N_\mathrm{Tot}$} &
\colhead{$N_{200}$} & 
\colhead{$s$} \\
\colhead{} & 
\colhead{(deg)} & 
\colhead{(deg)} & 
\colhead{} & 
\colhead{(km s$^{-1}$)} & 
\colhead{(Mpc)} & 
\colhead{$(10^{14} M_{\odot})$} & 
\colhead{} &
\colhead{} & 
\colhead{}
}

\startdata
NGC3054 & 148.619110 & -25.703440 & 0.007700 & 232 & 0.443 & 0.100 & 134 & 3 & --- \\
Antlia & 157.505626 & -35.342200 & 0.009107 & 646 & 1.330 & 2.702 & 269 & 68 & 0.0078 \\
NGC3250 & 156.741641 & -39.906827 & 0.009213 & 116 & 0.254 & 0.019 & 168 & 5 & 0.0095 \\
NGC3258 & 157.358909 & -35.497493 & 0.009307 & 640 & 1.320 & 2.640 & 262 & 67 & 0.0077 \\
NGC3256 & 157.277321 & -44.070510 & 0.009491 & 145 & 0.322 & 0.038 & 8 & 7 & --- \\
NGC3347 & 160.551132 & -36.379851 & 0.009750 & 168 & 0.368 & 0.057 & 121 & 7 & 0.0025 \\
NGC3091 & 150.094578 & -19.636829 & 0.012261 & 263 & 0.575 & 0.219 & 216 & 16 & 0.0014 \\
Hydra & 159.196746 & -27.517746 & 0.012635 & 686 & 1.409 & 3.224 & 533 & 265 & 0.0020 \\
Abell 194 & 21.420000 & -1.407220 & 0.018000 & 460 & 0.974 & 1.071 & 1382 & 161 & 0.0032 \\
MKW4 & 180.990540 & 1.888290 & 0.020000 & 536 & 1.119 & 1.625 & 1001 & 136 & 0.0057 \\
Abell 639 & 160.054114 & -46.269091 & 0.020610 & 465 & 0.976 & 1.079 & 46 & 43 & 0.0029 \\
MKW1 & 150.133700 & -2.957300 & 0.021000 & 356 & 0.766 & 0.521 & 630 & 53 & 0.0007 \\
{[}YMV2007{]}7604 & 36.527960 & -0.331920 & 0.021311 & 135 & 0.296 & 0.030 & 211 & 6 & --- \\
HCG97 & 356.845580 & -2.325990 & 0.021800 & 460 & 0.962 & 1.036 & 28 & 28 & --- \\
WBL081 & 39.571697 & 2.082272 & 0.021920 & 447 & 0.945 & 0.981 & 670 & 81 & 0.0029 \\
IC1860 & 42.403800 & -31.188600 & 0.022300 & 438 & 0.923 & 0.915 & 151 & 38 & 0.0036 \\
Abell 3581 & 211.887636 & -27.064399 & 0.022409 & 448 & 0.938 & 0.960 & 212 & 24 & 0.0038 \\
WBL074 & 37.911338 & 1.286248 & 0.022726 & 391 & 0.836 & 0.680 & 871 & 90 & 0.0036 \\
Abell 2870 & 16.937456 & -46.924795 & 0.023216 & 273 & 0.594 & 0.244 & 112 & 22 & 0.0022 \\
NGC1132 & 43.236044 & -1.236168 & 0.023348 & 221 & 0.487 & 0.134 & 394 & 15 & 0.0039 \\
Abell 2877 & 17.474568 & -45.920728 & 0.023874 & 705 & 1.434 & 3.433 & 227 & 104 & 0.0022 \\
Abell 3389 & 95.535893 & -64.956146 & 0.026615 & 524 & 1.082 & 1.478 & 35 & 30 & --- \\
MKW8 & 220.159160 & 3.476390 & 0.027000 & 521 & 1.087 & 1.501 & 1882 & 165 & 0.0011 \\
{[}YMV2007{]}2116 & 46.385427 & -0.378560 & 0.027619 & 504 & 1.050 & 1.352 & 709 & 59 & -0.0065 \\
Abell 4049 & 357.946318 & -28.359145 & 0.028116 & 315 & 0.679 & 0.366 & 338 & 28 & 0.0021 \\
Abell 4038 & 356.929990 & -28.141390 & 0.028200 & 741 & 1.499 & 3.935 & 574 & 119 & 0.0022 \\
Abell 3744 & 316.801240 & -25.438060 & 0.038100 & 588 & 1.205 & 2.063 & 117 & 66 & 0.0018 \\
Abell 3733 & 315.418321 & -28.042699 & 0.038980 & 593 & 1.213 & 2.108 & 96 & 72 & -0.0008 \\
WBL288 & 162.426979 & 0.337996 & 0.039426 & 322 & 0.692 & 0.391 & 851 & 43 & -0.0017 \\
CL1058+0137 & 164.566750 & 1.623722 & 0.039555 & 381 & 0.810 & 0.627 & 1233 & 105 & --- \\
Abell 463 & 67.307578 & -53.825798 & 0.040085 & 628 & 1.276 & 2.458 & 170 & 60 & 0.0009 \\
Abell 957 & 153.414132 & -0.911491 & 0.044159 & 655 & 1.330 & 2.789 & 1719 & 150 & 0.0029 \\
Abell 119 & 14.076250 & -1.216670 & 0.044200 & 859 & 1.706 & 5.888 & 1099 & 277 & 0.0020 \\
Abell 168 & 18.744541 & 0.411680 & 0.044841 & 523 & 1.081 & 1.501 & 1135 & 151 & 0.0019 \\
MS0116.3-0115 & 19.723390 & -1.003910 & 0.045200 & 438 & 0.917 & 0.916 & 866 & 55 & 0.0020 \\
Abell 3716 & 312.880616 & -52.687980 & 0.045716 & 762 & 1.526 & 4.221 & 464 & 205 & --- \\
Abell 1631 & 193.220092 & -15.420825 & 0.046422 & 652 & 1.324 & 2.757 & 639 & 204 & 0.0002 \\
{[}YMV2007{]}7609 & 17.096840 & 0.089780 & 0.046880 & 43 & 0.319 & 0.001 & 14 & 5 & 0.0009 \\
Abell 1644 & 194.290420 & -17.400280 & 0.047300 & 808 & 1.608 & 4.952 & 654 & 189 & -0.0004 \\
MCXCJ1121.7+0249 & 170.424379 & 2.814099 & 0.048607 & 575 & 1.177 & 1.943 & 574 & 134 & --- \\
Abell 4059 & 359.259580 & -34.760560 & 0.048700 & 791 & 1.577 & 4.674 & 454 & 174 & 0.0072 \\
Abell 2717 & 0.800420 & -35.927220 & 0.049000 & 576 & 1.177 & 1.943 & 426 & 81 & 0.0067 \\
IC1365 & 318.479955 & 2.569276 & 0.049264 & 662 & 1.339 & 2.859 & 849 & 153 & --- \\
MCXCJ0413.9-3805 & 63.502253 & -38.110323 & 0.049812 & 470 & 0.969 & 1.086 & 158 & 31 & --- \\
Abell 620 & 147.250000 & -26.033500 & 0.049900 & 544 & 1.025 & 1.283 & 5 & 5 & --- \\
{[}YMV2007{]}841 & 16.084398 & -0.239089 & 0.050225 & 247 & 0.529 & 0.177 & 85 & 16 & -0.0063 \\
RXJ2137.1+0026 & 324.277920 & 0.447500 & 0.051000 & 316 & 0.671 & 0.361 & 358 & 24 & 0.0074 \\
{[}YMV2007{]}4713 & 214.349900 & 0.088181 & 0.052894 & 251 & 0.538 & 0.186 & 955 & 13 & -0.0003 \\
Abell 151 & 17.208750 & -15.410000 & 0.053300 & 716 & 1.436 & 3.546 & 489 & 182 & 0.0005 \\
MKW6 & 214.402700 & 2.061870 & 0.053670 & 588 & 1.199 & 2.064 & 1231 & 137 & 0.0021 \\
Abell 978 & 155.120470 & -6.511394 & 0.054137 & 595 & 1.212 & 2.131 & 897 & 128 & 0.0012 \\
Abell 993 & 155.462316 & -4.928168 & 0.054516 & 400 & 0.840 & 0.710 & 832 & 66 & -0.0012 \\
Abell 3225 & 62.370780 & -59.604333 & 0.054979 & 1050 & 2.028 & 10.004 & 50 & 49 & 0.0014 \\
{[}YMV2007{]}49 & 214.331510 & 2.344280 & 0.055120 & 305 & 0.633 & 0.303 & 932 & 11 & 0.0005 \\
{[}YMV2007{]}514 & 24.315452 & -0.485475 & 0.055666 & 189 & 0.416 & 0.086 & 164 & 16 & 0.0012 \\
Abell 133 & 15.675420 & -21.873610 & 0.056600 & 667 & 1.317 & 2.743 & 41 & 17 & 0.0026 \\
ESO351-021 & 13.755800 & -35.319479 & 0.056897 & 383 & 0.763 & 0.533 & 17 & 7 & -0.0025 \\
Abell 2457 & 338.919980 & 1.484890 & 0.057800 & 684 & 1.376 & 3.129 & 1227 & 216 & 0.0004 \\
Abell 3880 & 336.971154 & -30.589133 & 0.057983 & 642 & 1.295 & 2.613 & 471 & 113 & -0.0043 \\
Abell 2415 & 331.418760 & -5.593330 & 0.058100 & 684 & 1.374 & 3.117 & 761 & 155 & 0.0014 \\
Abell 970 & 154.347500 & -10.677500 & 0.058700 & 777 & 1.543 & 4.420 & 179 & 142 & 0.0001 \\
Abell 3158 & 55.739224 & -53.642113 & 0.059279 & 1024 & 1.989 & 9.469 & 342 & 307 & 0.0009 \\
Abell 3223 & 62.078512 & -30.902245 & 0.059778 & 589 & 1.194 & 2.048 & 181 & 60 & 0.0000 \\
Abell 3266 & 67.842273 & -61.448160 & 0.059788 & 1275 & 2.431 & 17.300 & 536 & 439 & --- \\
Abell 3128 & 52.479170 & -52.566670 & 0.059900 & 825 & 1.633 & 5.241 & 815 & 374 & 0.0001 \\
CODEX-59603 & 359.419804 & 0.798581 & 0.061004 & 256 & 0.543 & 0.193 & 195 & 13 & 0.0013 \\
Abell 3809 & 326.740840 & -43.910000 & 0.062300 & 541 & 1.107 & 1.637 & 264 & 117 & -0.0020 \\
Abell 2734 & 2.836250 & -28.855000 & 0.062500 & 629 & 1.268 & 2.464 & 680 & 97 & -0.0015 \\
Abell 3135 & 53.509350 & -39.000150 & 0.064200 & 719 & 1.426 & 3.507 & 190 & 49 & -0.0014 \\
Abell 3122 & 50.576250 & -41.338890 & 0.064300 & 736 & 1.462 & 3.776 & 129 & 80 & -0.0027 \\
Abell 1069 & 159.935000 & -8.683610 & 0.065000 & 684 & 1.369 & 3.105 & 478 & 101 & -0.0053 \\
MCXCJ0340-4542 & 55.192363 & -45.682315 & 0.068939 & 349 & 0.715 & 0.444 & 19 & 11 & -0.0050 \\
Abell 3104 & 48.582500 & -45.424170 & 0.073000 & 362 & 0.699 & 0.417 & 36 & 5 & -0.0033 \\
Abell 3112 & 49.509398 & -44.251381 & 0.073973 & 398 & 0.821 & 0.675 & 59 & 23 & -0.0085 \\
MCXCJ0229.3-3332 & 37.337455 & -33.503170 & 0.076514 & 677 & 1.342 & 2.957 & 213 & 48 & -0.0003 \\
WHLJ012023-000444 & 20.096410 & -0.078980 & 0.078000 & 360 & 0.747 & 0.510 & 337 & 23 & 0.0037 \\
{[}YMV2007{]}3820 & 201.835649 & 1.276449 & 0.080432 & 498 & 1.011 & 1.270 & 901 & 48 & --- \\
MCXCJ1326.2+0013 & 201.573300 & 0.225100 & 0.082600 & 523 & 1.059 & 1.462 & 984 & 59 & 0.0198 \\
Abell 1750 & 202.801540 & -1.714170 & 0.083700 & 366 & 0.759 & 0.539 & 565 & 33 & --- \\
{[}YMV2007{]}4832 & 149.976097 & -0.234508 & 0.090357 & 421 & 0.849 & 0.759 & 554 & 17 & 0.0001 \\
Abell 2440 & 335.970430 & -1.637780 & 0.090600 & 811 & 1.580 & 4.892 & 479 & 135 & 0.0217 \\
{[}YMV2007{]}1540 & 154.946730 & -0.646272 & 0.093247 & 421 & 0.858 & 0.786 & 645 & 30 & 0.0150 \\
Abell 954 & 153.436363 & -0.121503 & 0.094641 & 746 & 1.459 & 3.872 & 1002 & 95 & -0.0024 \\
\enddata
\tablecomments{This table is published in the electronic edition of the {\it Astrophysical Journal} and is also available in various digital formats in the project's website (\url{https://splus-scale.github.io}).}
\end{deluxetable*}

\section{S-PLUS photo-z member selection}
\label{sec:ErikANDVitor-Section}

Cluster membership is best established from spectroscopic redshifts, for example, by using the caustic technique \citep[e.g.][]{Kang+2024}, where galaxies in projected phase space form a trumpet-shaped distribution bounded by caustics that separate members from interlopers. In this work, we adopted the shifting-gapper technique (see Section \ref{sec:Sample}). However, spectroscopic redshifts are often unavailable, particularly for faint galaxies lacking emission lines. In such cases, photometric membership determination has become essential for galaxy evolution and large-scale structure studies.

The 12 photometric bands of S-PLUS enable photo-$z$ estimates with relatively small errors compared to 5-band photometry, using machine-learning techniques \citep{Lima+22}. In Section \ref{sec:photo-z-performance} we present the achieved precision of S-PLUS photo-$z$s as a function of magnitude and redshift, based on spectroscopically confirmed galaxies. A more detailed and complete assessment of the accuracy of the S-PLUS photo-$z$s is provided by \citet{Lima+22}. In Section \ref{sec:photo-z-spec-z-comparison}, we compare photo-$z$ and spectroscopic redshift distributions for 10 clusters spanning $0.02 \leq z \leq 0.1$, and in Section \ref{sec:photo-z member-Section} we provide the selection of photometric members for Abell\,4038 using the Reliable Photometric Membership (RPM) algorithm, developed by \cite{lopes2020reliable}.

\subsection{Photometric redshift  performance} \label{sec:photo-z-performance}

A crucial step before employing photo-$z$s for membership determination is to assess their performance and identify potential limitations. 

As reported in \citet{Lima+22}, the performance of single-point estimates (SPEs) is evaluated using three metrics: the normalized median absolute deviation \citep[$\sigma_{\text{NMAD}}$,][]{Brammer2008}, which quantifies the scatter of the predictions; the bias, $\delta z$, which measures systematic over- or under-estimation of the photo-$z$; and the fraction of outliers, $\eta$, as defined by \citet{Ilbert2006}. For the probability density functions (PDFs), another three metrics are analyzed: the odds \citep{Benitez2000}, the Probability Integral Transform (PIT, \citealt{Polsterer2016}), and the Highest Probability Density Credible Interval (HPDCI, \citealt{HPDCI}), all related to how well-calibrated the PDFs are with respect to the uncertainty of the predictions. 

\begin{figure*}
    \centering
    \includegraphics[width=1.0\linewidth]{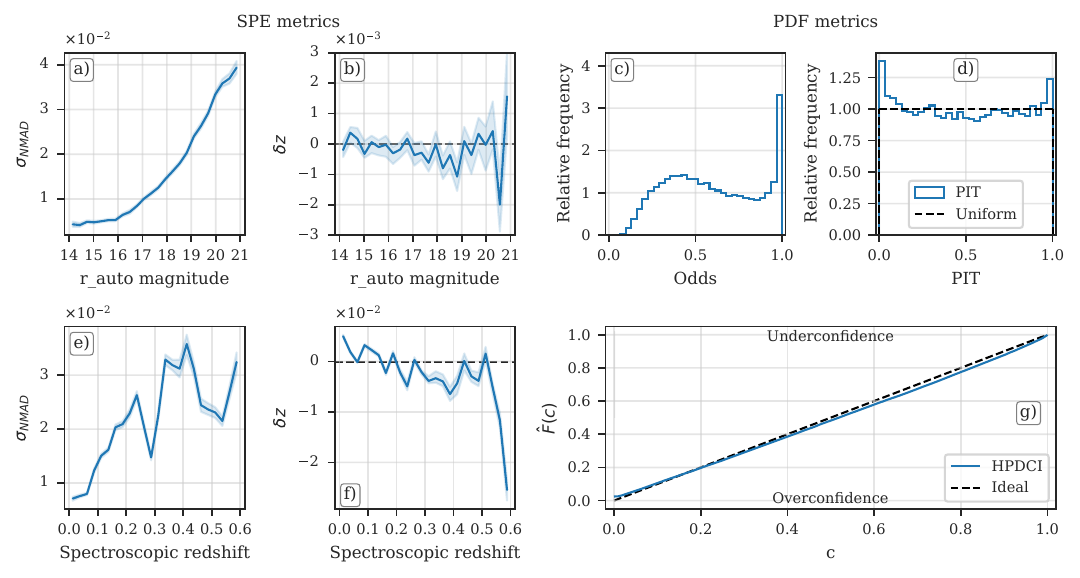}
    \caption{Performance of the photo-$z$ model. Panels (a) and (b) show the scatter and bias of the SPE metrics as a function of magnitude, and panels (e) and (f) as a function of spectroscopic redshift. Panels (c), (d), and (g) display the odds, PIT, and HPDCI metrics, respectively. Shaded regions represent $1\sigma$ uncertainties from bootstrap resampling.}
    \label{fig:photo-z-summary}
\end{figure*}

Figure \ref{fig:photo-z-summary} summarizes the performance of the photo-$z$s using the defined metrics except for the outlier fraction. Panels (a) and (b) show the SPE metrics as a function of \texttt{r\_auto} magnitude, and panels (e) and (f) as a function of spectroscopic redshift. As expected, the scatter increases for fainter objects and at higher redshifts as a result of larger photometric errors. The bias is negligible overall but increases for very nearby ($z \leq 0.05$) and distant ($z \geq 0.5$) objects, likely reflecting the under-representation of these regimes in the training sets. For the faint and nearby galaxy population, the predicted redshifts tend to be higher than the spectroscopic measurements, and for the distant galaxies, the predicted photo-zs are lower than the spec-zs. Panel (c) shows the odds distribution, with a peak near unity corresponding to bright galaxies (\texttt{r\_auto} $\leq 17$ mag) and lower values for fainter systems. The PIT histogram (panel d) and the HPDCI curve ($c$ vs $\hat{F}(c)$ in panel g, defined in \citealt{HPDCI_2} and where $c$ represents the threshold credibility) indicate well-calibrated PDFs, with no significant biases and only mild overconfidence.

The photo-z bias is quantified both as a function of magnitude and spectroscopic redshift, as shown in the panels b) and f) of Figure \ref{fig:photo-z-summary}, respectively. The bias as a function of magnitude is very low (up to $2\times10^{-3}$ for r $\approx$ 21). However, we see an increased bias as a function of redshift, especially for very nearby ($z \leq 0.05$) and distant ($z \geq 0.5$) objects. From our analysis, we can verify that the model has a bias for the faint+nearby galaxy population, where the predicted redshifts tend to be higher than the spectroscopic measurements, and for the distant galaxies, where the predicted photo-zs are lower than the spec-zs. This bias originates from the lack of training data in these regions of the parameter space, which are difficult to observe and obtain spectroscopic redshifts, and there is ongoing work with the goal of eliminating or diminishing its effects. 

In summary, we find precisions of $\sigma_\text{NMAD} \leqslant 1$\%, 2.2\% and 3.3\% for galaxies with $r \leqslant 17$, $19$ and $20$ mag, respectively, and $\sigma_\text{NMAD} \leqslant 2.2$\% and 3.2\% for galaxies at $z=0.3$ and $z=0.4$, with an overall outlier fraction of 1.6\%. 


\subsection{Comparison of spec-z and photo-z distributions for 10 clusters}

\label{sec:photo-z-spec-z-comparison}

\begin{figure*}[!ht]
    \centering
    \includegraphics[width=\linewidth]{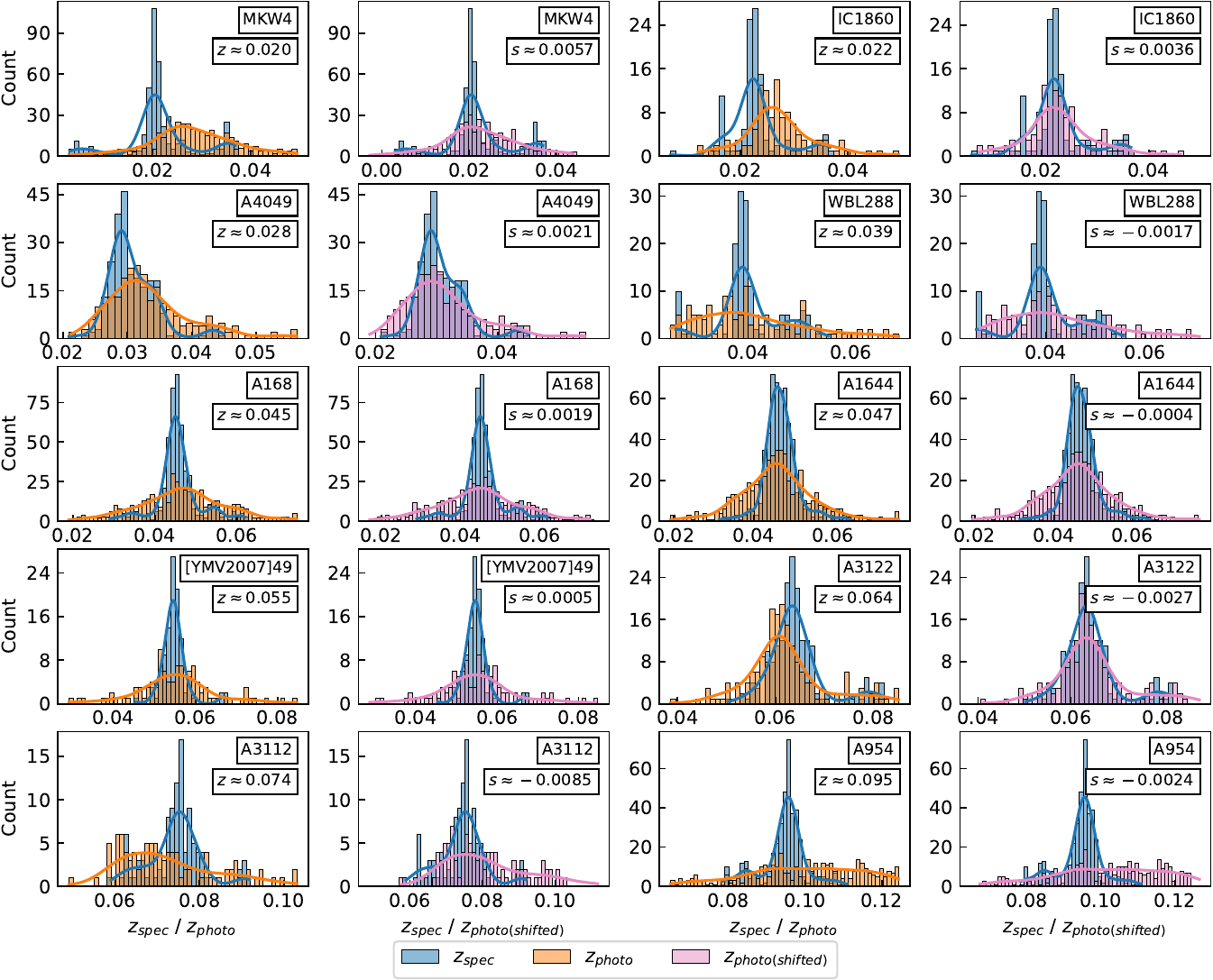}
    \caption{
    Comparative redshift distributions for 10 clusters with $0.02 < z < 0.1$. Blue histograms show spectroscopic members from \citet{Lima+24}. For each cluster, the left panel compares spectroscopic redshift (blue) and photo-$z$ (orange), while the right panel shows the corrected photo-$z$ distribution (pink) after applying the small shift $s$ (listed in the last column of Table \ref{tab:59clusters}) that aligns the peaks. Labels in the upper right of each panel list the cluster name, mean spectroscopic redshift ($z$), applied shift ($s$). Gaussian or double-Gaussian fits are over-plotted. After correction, the photo-$z$ distributions reproduce the spectroscopic redshift histograms well.}
    \label{fig:zoffset}
\end{figure*}

To assess the effectiveness of S-PLUS photo-$z$s in reproducing the spectroscopic redshift of cluster and group galaxies, in Figure \ref{fig:zoffset} we compare the redshift distributions for 10 systems in our sample, chosen to cover a range of redshifts. Each cluster is shown with two panels, with histograms constructed from members within $5R_{200}$ (see Section \ref{sec:Sample}). The spectroscopic redshifts of the cluster members taken from the literature compilation of \citet{Lima+24}, and listed in Table 2, are shown in blue and the corresponding photo-$z$ distributions for the same spectroscopic members (also listed in the same Table) appear in orange. Small offsets are present in most cases, but do not affect the overall agreement as the photo-$z$s consistently show well-defined peaks with low scatter. The offsets were estimated by fitting Gaussian functions (or double-Gaussian, when required) to the  spectroscopic and photometric distributions (blue and orange curves in Figure \ref{fig:zoffset}), and measuring the separation between their peaks. The labels in the upper-right portion of each panel indicate the cluster redshift ($z$), and the measured small shift ($s$) between the photo-$z$ and spectroscopic redshift distributions. Once corrected by the small offset $s$, the photo-$z$ distributions (pink curves) are brought onto the same reference frame as the spectroscopic redshift distributions, enabling joint analyses. 

In the last column of Table\,\ref{tab:59clusters}, we list the offsets between the photometric and spectroscopic distributions for the clusters and groups.
If the user wants to have spectroscopic redshift and photo-$z$ distributions matched for cluster member galaxies, the corresponding offset (of the order of 0.001) must be subtracted from the photo-$z$ values listed in this paper.

\subsection{Selecting new photometric members of A4038 using the RPM algorithm}
\label{sec:photo-z member-Section}

The Reliable Photometric Membership algorithm, developed by \cite{lopes2020reliable}, presents a machine learning–based approach that uses a support vector machine (SVM) to assign membership probabilities to galaxies based on their photometric properties. \cite{lopes2020reliable} demonstrated RPM's high accuracy in identifying spectroscopic members from photometric data in the local Universe ($z \leq 0.045$), achieving a completeness of $\sim 92\%$ and a purity of $\sim 87\%$. The method performs particularly well within $R_{200}$, reaching $\sim$95\% completeness and $\sim$94\% purity for $r < 0.5$ $R_{200}$. RPM facilitates studies of galaxy populations and reliable cluster richness estimates, proving effective across broad ranges of magnitudes and galaxy types.

As a practical application, we selected the galaxy cluster A4038 as our case study. Although relatively poorly studied, it is well covered by S-PLUS photometry. X-ray observations with XMM-Newton analyzed by \cite{Rossetti2010} indicate ongoing interactions and describe the cluster as thermodynamically controversial: despite lacking a pronounced central temperature drop, it exhibits a low central cooling time, leading to classifications as either cool core \citep[e.g.,][]{Peres1998} or non–cool core \citep[e.g.,][]{Sanderson2006}. Optical studies by \cite{Burgett2004} report large-scale substructure within 1.3 $R_{200}$ using 154 galaxy members, suggesting that A4038 is not dynamically relaxed. Its brightest cluster galaxy appears to be undergoing a merger and is surrounded by close companions, reinforcing the scenario of ongoing mass assembly in the core. Future analyses of diffuse intracluster light, such as those presented by \cite{LimaNeto2025} for NGC 5098/5096, further illustrate the diagnostic power of low-surface-brightness features for constraining cluster dynamical histories.

The RPM method was used to select photometric members of A4038 within $10 R_{200}$, using as a training sample the SCALE cluster/group galaxies listed in Table 2 (spectroscopic members selected through the shifting gapper technique), deliberately excluding A4038 galaxies to prevent calibration bias. The training stage utilized 15 features, including broad- and narrow-band colors (\texttt{u\_g\_auto}, \texttt{g\_r\_auto}, \texttt{r\_i\_auto}, \texttt{i\_z\_auto}, \texttt{g\_J0410\_auto}, \texttt{r\_J0660\_auto}, \texttt{J0410\_J0515\_auto}), magnitudes (\texttt{u\_auto}, \texttt{g\_auto}, \texttt{r\_auto}, \texttt{i\_auto}, \texttt{z\_auto}), photo-$z$s (\texttt{photo\_z}), surface brightness (\texttt{mu\_max\_g}), and projected distance to the cluster center (\texttt{dist\_mpc}) and the \texttt{flag\_member} identification, assuming that the spectroscopic members determined by the shifting-gapper method \citep{Fadda+96,Lopes+09} define the calibration sample.

Although the spectroscopic sample is complete only down to $r \approx 17.5$ mag, a substantial number of redshifts are available as faint as $r \approx 19$ mag (corresponding to $M_r \approx -16.2 + 5\log_{10} h$ mag at the cluster redshift). RPM is therefore expected to recover a significant population of faint photometric members missed by previous spectroscopic surveys, while also identifying a smaller number of additional bright members.

Using $R_{200} \approx 1.5~h^{-1}$ for A4038 (Section \ref{sec:Sample}), the prediction of membership within $5 R_{200}$ resulted in 434 photometric members, with 77\% completeness and 71\% purity, increasing to 96\% completeness and 93\% purity within $0.5R_{200}$. These values were obtained using a conservative threshold of membership probability greater than 70\%.  Performance in the central region is comparable to that found by \citet{lopes2020reliable} and declines towards larger radii, as expected, while remaining acceptable.

Our color analysis (Figure \ref{colorconc}; Section \ref{sec:color_segregation_by_lum_con}) for galaxies inside $5 R_{200}$ is based on 434 RPM-selected photometric members, of which 296 were previously identified as spectroscopic members (80\% of the 356 galaxies with \texttt{flagmember} = 0 and with S-PLUS data). We note that 63 galaxies classified as spectroscopic members lack S-PLUS data and therefore have no RPM classification. From the 434 RPM-selected members, we removed 14 galaxies whose spectroscopic redshifts identified them as interlopers, yielding a final sample of 420 galaxies. Overall, RPM identified 124 new photometric members within $5R_{200}$, most of them fainter than $r = 17.5$ mag.

For the analysis of the substructure (Figure \ref{calsagos_DSPlus}; Section \ref{sec:A4038-2d}), we use the full RPM photometric sample, including the 124 new photometric members within $5 R_{200}$ and 213 beyond this radius. The latter should be interpreted with caution due to larger uncertainties at greater clustercentric distances.

\section{Results} \label{sec:Andre-Daniela-Section}

In this section, we use the spectroscopic and photometric members to find substructures in Abell \,4038 using a 3D technique (using phase-space information) and a 2D technique (positions only). We then study the color segregation by luminosity and radius and by luminosity and galaxy concentration.

\subsection{Substructures around A4038  using 3D phase-space information} \label{sec:A4038-3d}

In this subsection, we perform a substructure analysis using positions and velocities of all spectroscopic member galaxies (selected via the shifting-gapper technique; see Section \ref{sec:Sample}) within a projected clustercentric distance of $10\ h^{-1}$ Mpc (419 inside $5 R_{200}$). We apply DS+, a modified version of the Dressler-Schectman test \citep{dressler1988evidence} improved by \cite{biviano2017concentration} and \cite{benavides2023ds+}. Unlike the standard version, DS+ does not adopt a fixed number of neighbors, but explores different multiplicities and evaluates kinematic deviations relative to the cluster. Significance is assessed via Monte Carlo resampling. Overlapping groups are removed and nearby systems in distance and velocity are merged to avoid fragmentation. Velocity dispersions larger than that of the cluster are not considered. DS+ is sensitive to spatially compact subsystems with distinct mean velocity and/or velocity dispersion (see \citealt{benavides2023ds+}).

We also use \texttt{mclust} \citep{fraley2002model}, which applies Gaussian Mixture Models (GMM; \citealt{Reynolds2009}) to determine the number of components and assign galaxy memberships with associated uncertainties \citep{scrucca2016mclust,scrucca2023model}. Model selection is performed using the Bayesian Information Criterion (BIC; \citealt{neath2012bayesian}), which penalizes overfitting and selects among 16 covariance parameterizations differing in shape, volume, and orientation.

The final substructure catalog combines DS+ and \texttt{mclust} results, retaining only statistically robust systems. For DS+, we require detections at the 90\% confidence level with at least 5 members. For \texttt{mclust}, we accept components whose scale parameters lie within $2\sigma$ of the mean, excluding excessively diffuse systems.

\begin{figure*}
    \centering
    \includegraphics[width=1\linewidth]{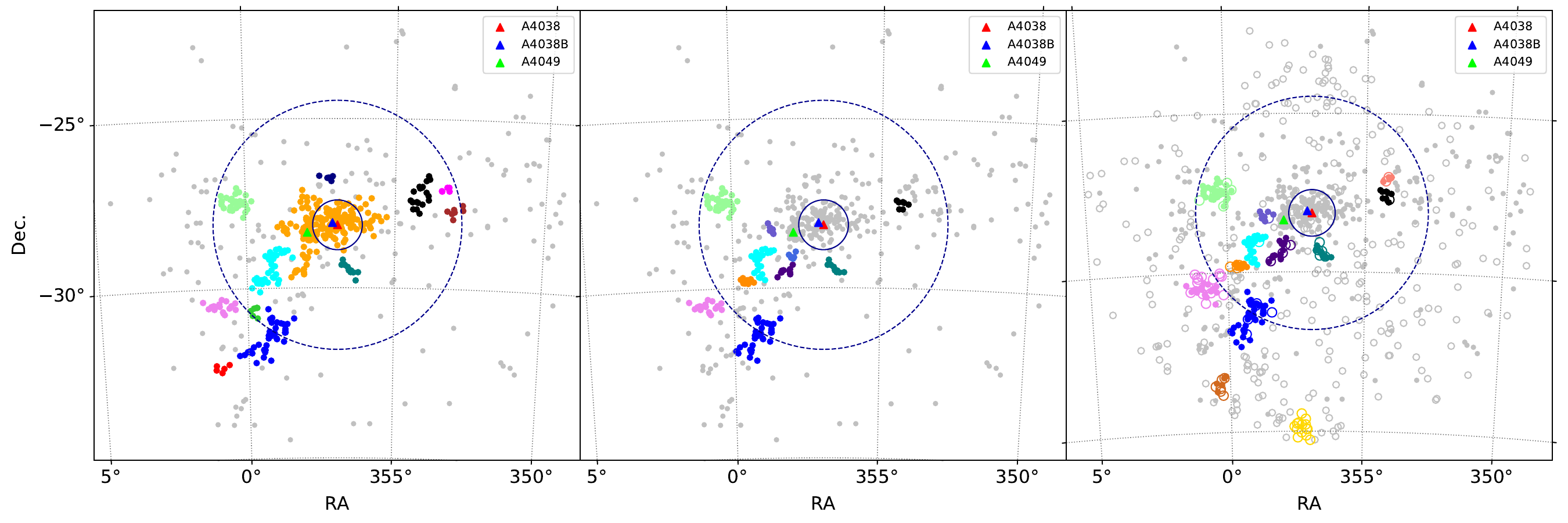}
\caption{Identification of galaxy substructures in the direction of A4038. The blue solid and dashed circles correspond to $R_{200}$ and $5 R_{200}$ of the cluster, respectively. {\it Left panel:} 3D phase-space substructures identified for A4038 using the spectroscopic sample and a combination of DS+ and \texttt{mclust} methods, as described in Section\,\ref{sec:A4038-3d}. {\it Middle panel}: 2D substructures recovered using \texttt{CALSAGOS} on the spectroscopic sample, as described in Section\,\ref{sec:A4038-2d}. {\it Right panel}: 2D substructure identification using \texttt{CALSAGOS} on the spectroscopic and photometric samples, as also described in Section\,\ref{sec:A4038-2d}. The filled circles correspond to spectroscopic cluster members and the open circles represent the photometric cluster members selected using the approach described in Section\,\ref{sec:photo-z member-Section}.}
    \label{calsagos_DSPlus}
\end{figure*}

The combination of DS+ and \texttt{mclust} identifies several subgroups in this region (left panel of Figure\,\ref{calsagos_DSPlus}). Four subgroups with more than 10 members are located within $5 R_{200}$ (excluding the central cluster), one of which may be part of A4038B (see Section \ref{core}), forming part of the infall region around A4038 and A4038B. Additional systems at larger radii trace the surrounding large-scale structure. In the next subsection, these results are compared with those obtained using the expanded sample.

\subsection{Substructures in A4038 using 2D information} \label{sec:A4038-2d}

We apply Clustering ALgorithmS Applied to Galaxies in Overdense Systems (\texttt{CALSAGOS}; \citealt{Olave-Rojas+23}) to identify substructures in and around A4038 using projected galaxy positions. We first analyze the spectroscopic members (Table 2) and then the combined spec-$z$ and photo-$z$ sample, where photo-$z$ members are galaxies without spectroscopy classified via the RPM method (Section\,\ref{sec:photo-z member-Section}). Unlike Section\,\ref{sec:A4038-3d}, only positional information is used, enabling the inclusion of photo-$z$ candidates and allowing comparison with the 3D phase-space analysis.

Substructures are identified with the 2D version of \texttt{LAGASU}, which implements DBSCAN (\citealt{1996Ester}). The required inputs are the cluster center, cluster redshift, $R_{200}$, the minimum number of galaxies defining a substructure ($N_{\rm mim}$), and the search radius $r_{\rm search}$ (see \citealt{Olave-Rojas+23}). 

To maintain consistency with Section \ref{sec:A4038-3d} while minimizing fragmentation and spurious detections due to projection effects, we adopt $N_{\rm mim}=5$ inside $5R_{200}$ and 10 outside. 
These thresholds were selected because they yielded the most robust configurations. To determine this, we perturbed the galaxy positions and reran the substructure search over 100 realizations. This procedure allowed us to assess the stability of the original detections by comparing them with those recovered from the perturbed samples. The tests showed that the most stable configurations correspond to systems with at least 5 members within $5R_{200}$ and 10 members beyond that, indicating that structures with these membership numbers are highly likely to be physically real.

The search radius $r_{\rm search}$ is defined as the median distance to the $K$-nearest neighbors of each galaxy. The same configuration is used for the combined sample. The result for the spectroscopic sample is shown in the middle panel of Figure\,\ref{calsagos_DSPlus}.

In the right panel of Figure\,\ref{calsagos_DSPlus}, photometric members are shown as open circles with the same colors as the corresponding spectroscopic substructures (filled circles). Cluster members are indicated by gray points. By construction, the 2D method does not identify substructures within the central region, as galaxies within $\sim R_{200}$ are assigned to the principal halo. 
Excluding this region, one also avoids the issues 
of chance alignments and unreal groupings in the most susceptible region.

To further evaluate the stability of the 2D identification in the spectroscopic sample of galaxies (selected within $10R_{200}$ via the shifting-gapper technique), we performed three statistical tests: (i) randomly removing 20\% of galaxies, (ii) randomly adding 80\%, and (iii) bootstrapping galaxy positions. Each procedure is repeated 100 times, and substructures are re-identified in each realization. We compare the original classification with the simulated samples using the Adjusted Rand Index (ARI) and the Jaccard index. The ARI values are 0.91, 0.99, and 0.61, while the mean Jaccard values are 0.2, 0.9, and 0.3, respectively, for each statistical test. These results indicate that the substructures are globally stable and are unlikely to arise from statistical fluctuations, although some individual systems are not recovered by resampling. Identification remains stable when photo-$z$ members are included, as shown in the right panel of Figure\,\ref{calsagos_DSPlus}.

\begin{figure}[ht]
    \centering
    \includegraphics[width=\linewidth]{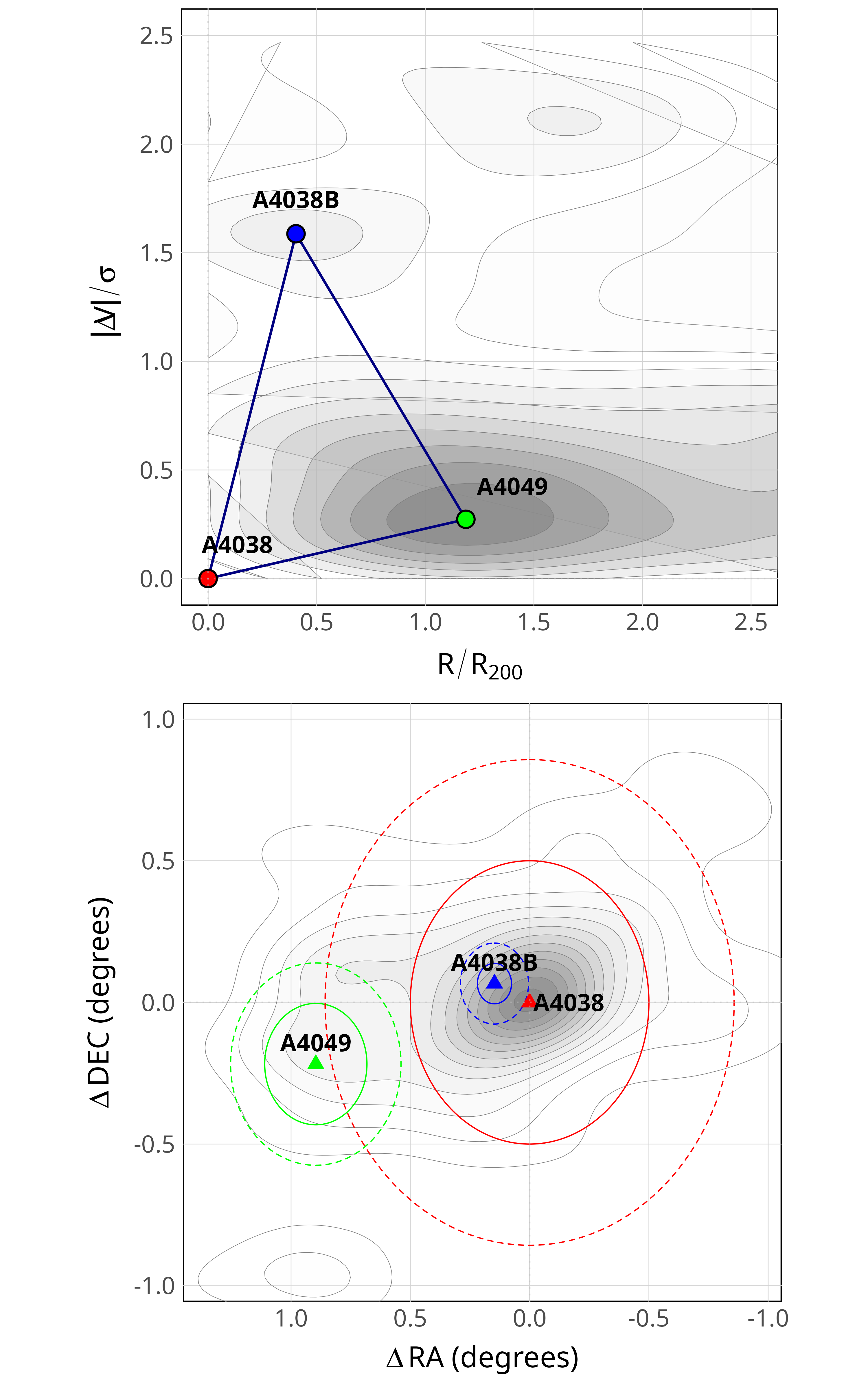}
    \caption{{\it Upper panel:} Density contours in projected phase space around A4038
showing the 2D density distribution of all galaxies. Colored points and labels mark the three cluster centers identified by \texttt{mclust} (+BCGs) and \texttt{pdfCluster}. The overdensity located at $|\Delta V|/\sigma = 2.2$ and $R/R_{200} = 1.7$ is most probably a substructure around A4038B. 
{\it Lower panel:} Spatial distribution in Equatorial coordinates, with density contours plus confidence ellipses at the 68\% (solid) and 95\% (dashed) levels. Triangles indicate the cluster centers.}
    \label{central}
\end{figure}

\subsection{Substructures in the core of A4038} \label{sec:central} 
\label{core}

Given the apparent complex structure of the central region of A4038 (the orange points in the left panel of Figure~\ref{calsagos_DSPlus}), we further investigate this region using position and velocity information and the techniques \texttt{mclust}  and \texttt{pdfCluster} (density-based clustering). The \texttt{mclust} analysis was performed with input of the coordinates of the 
BCGs previously identified from the imaging data of this region, thereby incorporating this structural information into the clustering solution.
Using clustercentric distances and velocity offsets, we assess whether the large central cluster contains smaller embedded substructures and kinematically distinct groups.
The latter method performs a density-based clustering through a nonparametric kernel density estimation, identifying clusters as connected high-density regions and attraction basins around density modes, thus accommodating irregular, non-Gaussian shapes and partial overlaps \citep{azzalini2014clustering}. Used together, these methods can uncover substructures that DS+ and \texttt{mclust} may miss, because the density-based step relaxes Gaussian assumptions and the need for strong local kinematic contrasts, capturing irregular, low-contrast or non-elliptical components. At the same time, the mixture model provides probabilistic memberships. 

Using this joint strategy on the central system, we find three structures whose centroids roughly coincide with the nominal centers of the A4038 and A4049 clusters \citep{hudson2001streaming}, plus a smaller system that we designate as A4038B, as shown in Figure \ref{central}.  The dynamical configuration of the A4038 system, as illustrated in the upper panel of this figure, reveals a complex interplay between its main components: A4038, A4038B (at position RA =  357.097066 deg and Dec = --28.074826 deg), and A4049 (at position RA = 356.929990 deg, Dec = --28.141390 deg). In particular, A4038B, while spatially close to A4038 in projection (slightly northwest and at low $R/R_{200}$ in phase-space), exhibits a significantly higher normalized absolute velocity ($|\Delta V|/\sigma \approx 1.5-2.0$), with a large line-of-sight velocity offset of $\approx 1.76-1.78 \times 10^3$ km s$^{-1}$ that suggests it is a background system. Meanwhile, A4049 (green symbol) is situated at a noticeably larger normalized radial distance ($R/R_{200} \approx 1.25$) but maintains a relatively low normalized absolute velocity ($|\Delta V|/\sigma < 0.5$), consistent with a very small line-of-sight velocity difference of $\approx 25$ km s$^{-1}$ relative to A4038. This combination, coupled with its clear spatial separation to the west/southwest of A4038, suggests a dynamical connection or ongoing merger process with A4038 where both may belong to the same large-scale structure. The overall phase-space and spatial distributions underscore the complex dynamical state of the A4038 system, indicative of significant substructures and potential interaction events. 

These findings point to a significant dynamical complexity in this large-scale system, which will be discussed in a forthcoming work.

\subsection{Color segregation by luminosity and radius for A4038}
\label{sec:color-segregation} 

We examine $g-r$ color distributions of photometric members in different absolute magnitude bins and as a function of projected distance from the center of the cluster to $5 R_{200}$. We derived absolute magnitudes and colors, taking into account the distance modulus (using the cosmology defined in Section \ref{sec:intro}) and extinction corrections with the values found in the extinction value-added catalog of S-PLUS DR5, columns \texttt{A\_r\_SFD} and \texttt{A\_g\_SFD}.

\begin{figure*}
    \centering
   \includegraphics[width=1\linewidth]{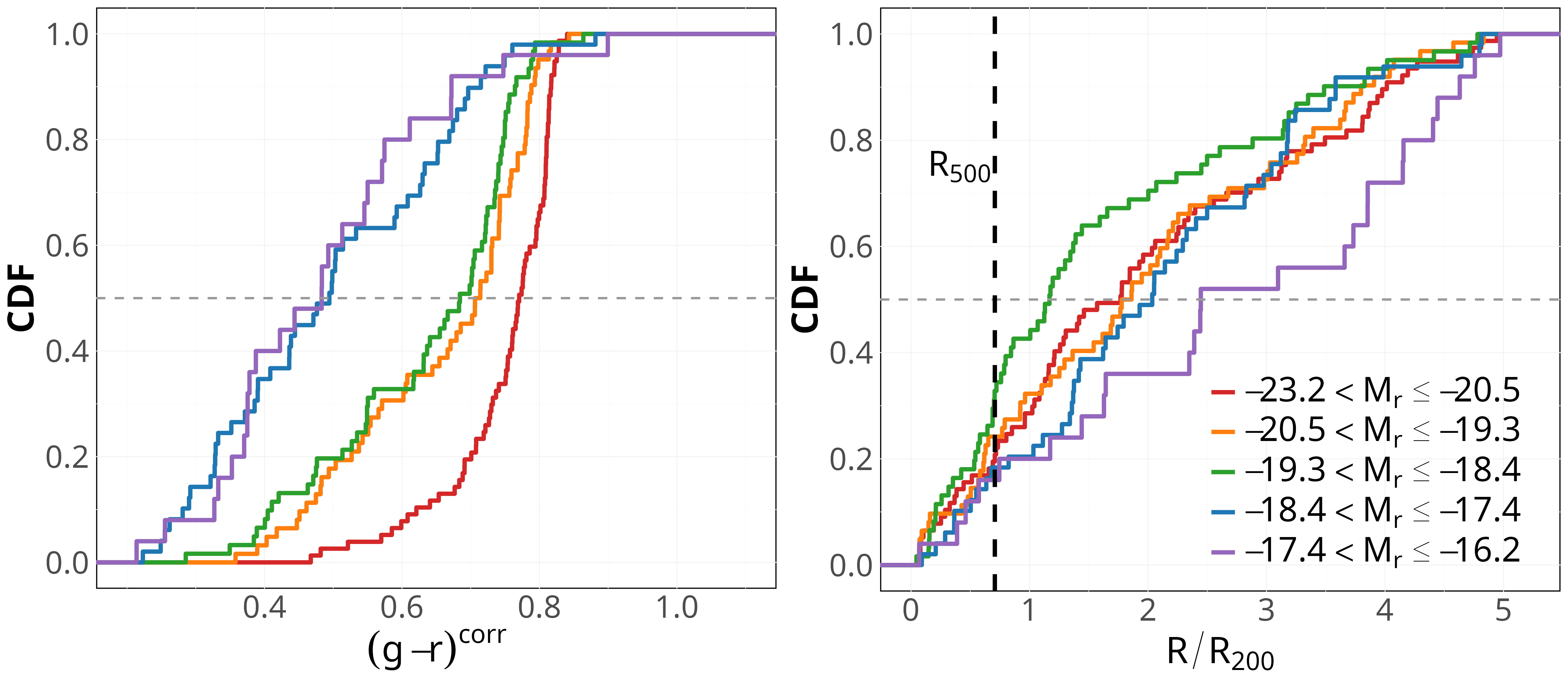}
   \caption{Cumulative distribution functions (CDFs) of the g − r color (left panel) and the normalized projected radial distance, $R/R_{200}$ (right panel), for galaxies grouped into five absolute magnitude quintiles: Q1 (brightest; red), Q2 (orange), Q3 (green), Q4 (blue), and Q5 (faintest; purple). The left panel illustrates segregation in (g-r) color (in units of mag) in bins of magnitude, while the right panel investigates the radial dependence of these magnitude-defined galaxy populations.}
 \label{cdf}
\end{figure*}

The magnitude bins were defined according to the quintiles of the distribution. The sample comprises 420 galaxies divided into five quintiles of absolute magnitudes of approximately 84 galaxies each: Q1 (brightest 20\%, $-23.2 + 5\log_{10} h \leq M_r \leq -20.5 + 5\log_{10} h$), Q2 ($-20.5 + 5\log_{10} h  < M_r \leq -19.3 + 5\log_{10} h$), Q3 ($-19.3 + 5\log_{10} h < M_r \leq -18.4 + 5\log_{10} h$), Q4 ($-18.4 + 5\log_{10} h < M_r \leq -17.4 + 5\log_{10} h$), and Q5 (faintest 20\%, $ -17.4 + 5\log_{10} h < M_r \leq -16.2$). This comparison is shown in Figure \ref{cdf}.

We note that no k-corrections were applied to the colors used here.
At the mean redshift of A4038 ($z=0.028$), the k-correction for $g-r$ is approximately 0.02~mag
\citep{chilingarian2012universal}. This shift is small compared to the full color range of our sample
($\gtrsim0.5$~mag in each bin) and does not affect the relative ordering of galaxies into quintiles or the
qualitative trends we discuss. However, for absolute color comparisons with other clusters at
different redshifts, k-corrections will be applied in future studies.

Given typical photometric uncertainties of $\sim 0.02$~mag, individual galaxies near quintile
boundaries may be misassigned. However, the mean colors of each quintile remain robust because
misclassification is symmetric and affects only a small fraction of the sample. A bootstrap test
(adding $0.02$~mag Gaussian noise) shows that quintile boundaries vary by less than $0.03$~mag,
confirming the stability of the derived trends.

The left panel reveals strong statistical evidence for color segregation across magnitude quintiles. 
An analysis of variance indicates that magnitude explains 35.7\% of color variance ($p < 2 \times 10^{-16}$). 
The brightest galaxies (Q1) exhibit the reddest colors with a median $g-r = 0.77$ mag, followed systematically by Q2 ($g-r = 0.71$ mag), Q3 ($g-r = 0.64$ mag), Q4 ($g-r = 0.47$ mag), and the faintest population Q5 ($g-r = 0.46$ mag). This systematic progression spans $\Delta(g-r) = 0.31$ mag between extreme quintiles, with the overall color-magnitude correlation reaching $r = -0.606$ ($p < 2 \times 10^{-16}$).

Pairwise Kolmogorov-Smirnov (KS) tests 
with Bonferroni correction\footnote{The Bonferroni correction is a statistical tool used to adjust the significance level when performing multiple statistical comparisons.}
reveal that 8 out of 10 comparisons (80\%) remain statistically significant, demonstrating robust color-magnitude relationships. The most pronounced separations occur between extreme quintiles: Q1 vs Q5 ($D = 0.770$, $p < 0.001$; large effect), Q1 vs Q4 ($D = 0.699$, $p < 0.001$; large effect), and Q1 vs Q3 ($D = 0.486$, $p < 0.001$; medium effect). Even intermediate comparisons show substantial effects: Q2 vs Q4 ($D = 0.464$, $p < 0.001$; medium effect) and Q3 vs Q5 ($D = 0.437$, $p < 0.001$; medium effect). Only adjacent quintiles show a non-significant separation after correction: Q2 vs Q3 ($D = 0.181$, $p = 1.000$) and Q4 vs Q5 ($D = 0.130$, $p = 1.000$).

In contrast, the right panel of Figure \ref{cdf} displays the cumulative distribution functions of the normalized projected radial distance ($R/R_{200}$) for these same magnitude quintiles. It shows that the radial distributions do not exhibit statistically significant segregation after multiple comparison corrections. This lack of significant radial segregation among populations grouped by magnitude is consistent with the negligible overall magnitude-radius correlation found ($r = 0.076$, $p = 0.208$). The statistical analysis also reveals only modest radial variations ---$Q1$ (median $R/R_{200} = 1.77$), $Q2$ ($1.82$), $Q3$ ($1.17$), $Q4$ ($2.04$), and $Q5$ ($2.45$)--- with none of the pairwise Kolmogorov-Smirnov comparisons appearing to be significant.

The largest radial differences are found between $Q3$ and $Q5$ ($D = 0.383$, $p_{\text{corrected}} = 0.076$) and between $Q3$ and $Q4$ ($D = 0.313$, $p_{\text{corrected}} = 0.075$), both corresponding to medium effect sizes but not reaching the statistical significance of
0.05. However, at the 90\% confidence level, the radial difference between $Q3$ and $Q5$ becomes statistically significant, indicating that galaxies in the faintest bin ($M_r > -17.4$) exhibit a more extended radial distribution than intermediate-luminosity galaxies ($-19.3 < M_r \leq -18.4$). Beyond the KS test, the Mann--Whitney U (Wilcoxon Rank-Sum) test was applied, revealing significant differences (at 95\% confidence) between Q3, Q5, and the combined Q1+Q2+Q4 quintiles, suggesting distinct radial distributions among these samples.

\begin{figure}
    \centering
    \includegraphics[width=0.93\linewidth]{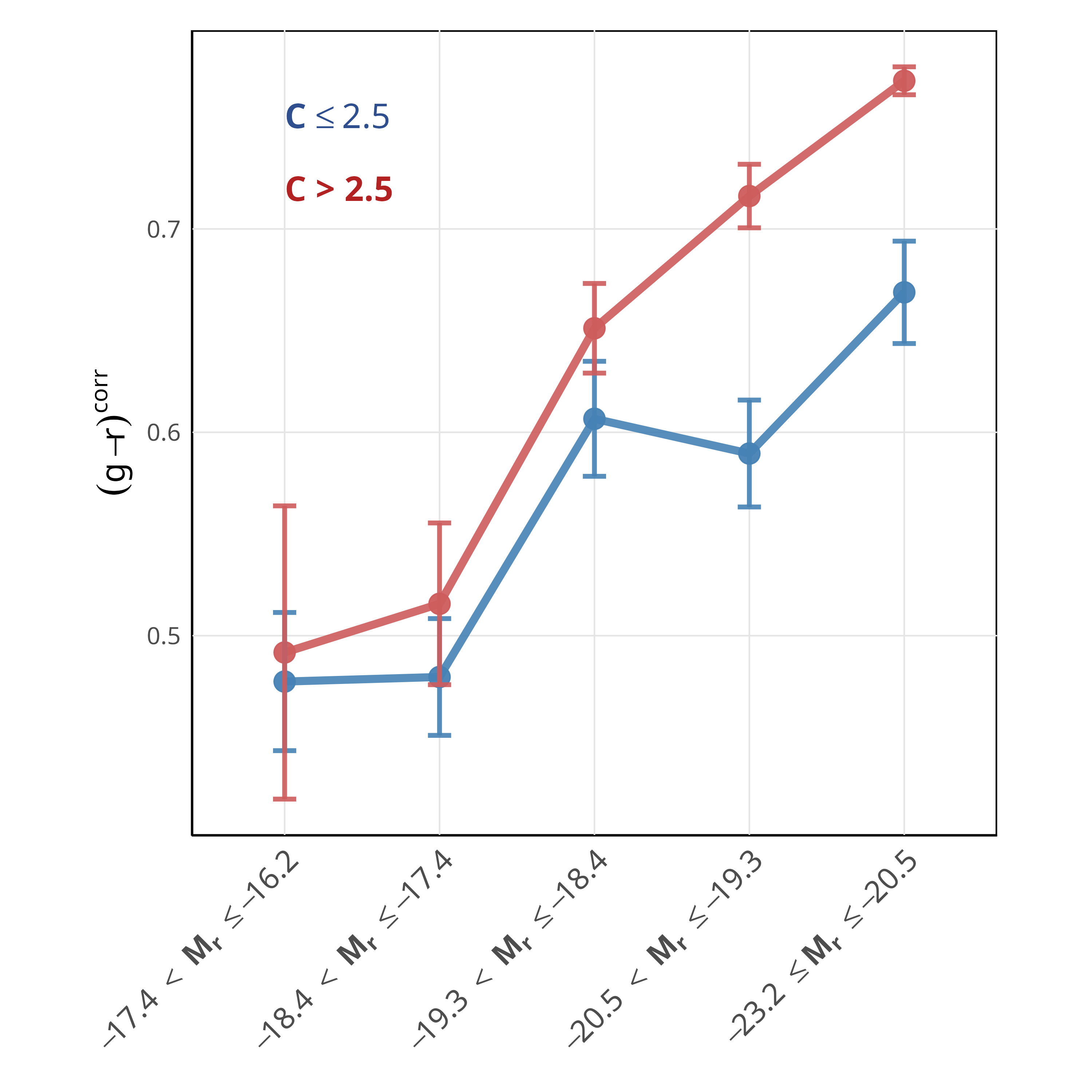}
    \vspace{0.25cm}
    \includegraphics[width=0.84\linewidth]{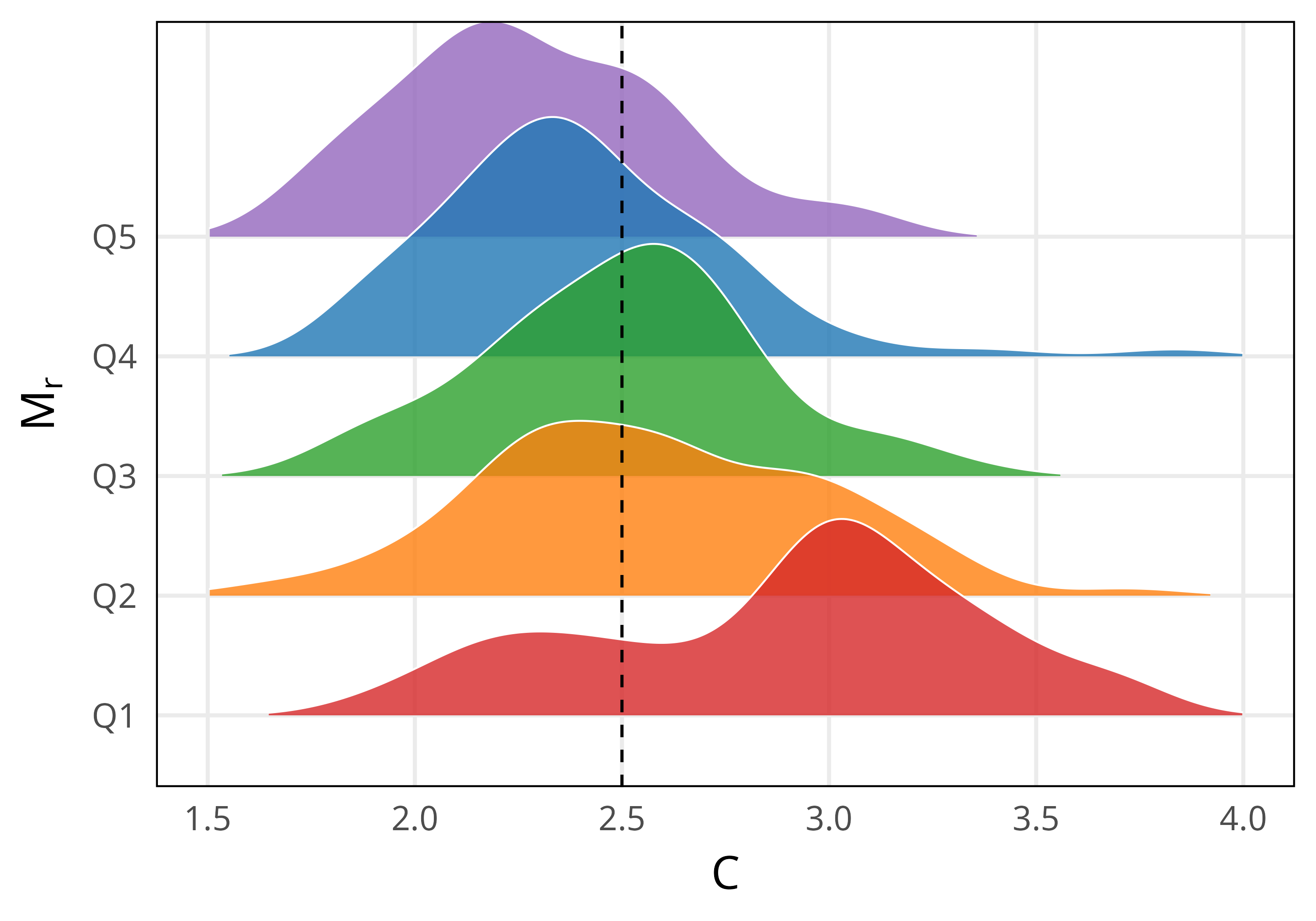}
    \caption{{\it Upper panel:} Behavior of $g-r$ color (in units of magnitude) as a function of absolute {\it r}-band magnitude ($M_r$) for two galaxy populations divided by their concentration index ($C$). Galaxies with $C \leq 2.5$ (blue line) are compared to those with $C > 2.5$ (red line). {\it Lower panel:} Density distributions of $C$ within each bin of absolute r magnitude. The colors are the same as in Figure \ref{cdf}.}
    \label{colorconc}
\end{figure}

\subsection{Color segregation by luminosity and concentration for A4038}
\label{sec:color_segregation_by_lum_con}

In the upper panel of Figure \ref{colorconc}, we present the behavior of the $g-r$ color as a function of the absolute magnitude in the {\it r} band ($M_r$). Galaxies are divided into two populations based on their concentration index, {\it C}, defined as the radius that contains 90\% of the light over the radius that contains 50\% of the light of the galaxy, $R_{90}/R_{50}$): those with $C \leq 2.5$ (blue line; 198 galaxies, 50.6\%) and those with $C > 2.5$ (red line; 193 galaxies, 49.4\%). In samples dominated by low-flux and low-surface brightness galaxies, higher concentration thresholds (e.g. $C\gtrsim 2.8$) are known to be biased by surface brightness limitations, which preferentially suppress the measured outer light profiles and artificially lower $R_{90}$. Early-type dwarf galaxies are intrinsically less concentrated than giant ellipticals, typically occupying the range $C \sim 2.3-2.6$. For this reason, we adopt a conservative threshold of $C = 2.5$, which preserves structural completeness while remaining physically consistent with the expected concentration of dwarf galaxies \citep{graham2003hst,blanton2005properties,janz2014near}. The magnitude bins shown on the x-axis of Figure \ref{colorconc} have 81 or 82 galaxies each. For the first (faintest; $-17.40 < M_r \leq -16.16$) and second ($-18.4 < M_r \leq -17.4$) bins, 28\% and 30\% of the galaxies, respectively, have $C > 2.5$, while for the brightest bin ($-23.17 \leq M_r \leq -20.50$), 24\% of galaxies have $C > 2.5$. The remaining two magnitude bins present more balanced distributions of $C$ (see the lower panel of Figure \ref{colorconc}). We plan to improve the statistics on these relations in the future by adding more clusters to this analysis; however, all bins in  Figure \ref{colorconc} contain a sufficient number of objects.

The two-way analysis of variance reveals that both magnitude quintile and concentration exert a statistically detectable influence on color. However, the effects of these on color are independent of each other. Our result indicates a general trend in which galaxy luminosity (or stellar mass) drives differences in their average color, regardless of whether they are concentrated or extended systems. 
However, galaxies with $C > 2.5$ show a significantly stronger color-magnitude correlation ($r = -0.631$, $p < 2 \times 10^{-16}$) compared to galaxies with $C \leq 2.5$ ($r = -0.431$, $p = 2.23 \times 10^{-10}$). Fisher's z-test confirms that these correlations are significantly different ($p = 0.0057$), indicating that the structure modulates the color-magnitude relationship. The analysis reveals systematic trends: Q1 shows 75.6\% of galaxies with $C > 2.5$ having a mean of $\langle C \rangle = 2.95$, while Q5 contains only 26.4\% of such systems with $\langle C \rangle = 2.33$. 
Our analysis also reveals that galaxy concentration is  correlated with both absolute magnitude and color. Specifically, we find a notable correlation between concentration and magnitude ($r = -0.487$, $p = 1.20 \times 10^{-24}$), and a very similar degree of correlation between concentration and color ($r = 0.436$, $p = 1.52 \times 10^{-19}$). Both correlations are statistically significant, confirming that brighter galaxies tend to have larger values of $C$ and redder colors, consistent with the expectations for the morphology-density relation in cluster environments.

Note that the concentration index $C$ for the faintest galaxies (Q5) could potentially be affected by the PSF, since galaxies below $1.5$~kpc are unresolved at $z=0.028$ (see FWHM $1.2''$). Nevertheless, all Q5 galaxies considered in this work have Petrosian radii between $5.3$ and $10.6$ arcsec ($\sim3.0$--$6.0$~kpc), far exceeding this limit. Hence, their $C$ values are reliable and not dominated by seeing effects.

\subsection{The convergence of concentration classes}

We should note that both concentration classes converge toward similar blue colors at the faint end ($M_r > -17.4$), where the color difference between extended and concentrated galaxies becomes negligible ($\Delta(g-r) \approx 0.02$ mag). This convergence indicates that faint galaxies are predominantly blue, regardless of their concentration index.

However, several observational considerations must be addressed when interpreting the convergence of concentration classes at faint magnitudes. Faint galaxies exhibit systematically lower peak surface brightness values (Q5: $\mu_{max,r}$ = 21.89 vs Q1: $\mu_{max,r}$ = 18.11 mag arcsec$^{-2}$, $t = -38.744$, $p = 1.61 \times 10^{-76}$), with correlations of the surface brightness with Petrosian radius ($r = -0.311$) and concentration ($r = -0.616$). The observed decrease in the Petrosian radius (Q1: 18.5 kpc vs Q5: 3.8 kpc) from bright to faint galaxies is consistent with the expected behavior of this metric for progressively lower surface-brightness galaxies. Although this pattern suggests that magnitude-limited samples may be biased against detecting extended systems among faint galaxies, visual inspection of our sample using the AstroInspect tool \citep{astroinspect} reveals the presence of both dwarf ellipticals (dE) and dwarf S0 (dS0) galaxies, rather than exclusively clumpy, low-surface-brightness systems. The apparent convergence to blue colors at faint magnitudes ($M_r > -17.4$) may then reflect a combination of observational selection effects and genuine astrophysical processes.
This finding complements broader discussions in the literature, where low-mass galaxies are often noted to follow different evolutionary paths compared to their more massive counterparts. For example, \cite{blanton2009physical} suggest that as luminosity decreases, galaxies tend to exhibit lower average concentrations, diminishing the ability of such structural parameters to distinctly separate populations based on their colors and magnitudes in this faint regime.
A detailed morphological analysis combining visual classification with surface photometry will be presented in a forthcoming paper to disentangle observational biases from intrinsic physical properties in the dwarf galaxy population of A4038.

\section{Discussion} \label{sec:discussion}

\subsection{On the comparison of the 2D and 3D phase-space techniques for substructure detection}

A comparison between the performance of the  3D  phase-space and 2D substructure identification methods can easily be seen in Figure \ref{calsagos_DSPlus}. In this figure, the left panel shows the 3D  phase-space identification using spectroscopic redshifts, the middle panel shows the 2D identification using spectroscopic redshifts, and the right panel shows the results using the 2D technique for an expanded sample that joins spectroscopic and photo-$z$ members. To further simplify the comparison, we have used the same color code in all the panels of the figure. 

In the area within $5 R_{200}$, four substructures with more than 10 members were detected using the 3D phase-space method, all recovered with \texttt{CALSAGOS} (2D technique) when using spectroscopic redshifts alone and when adding photo-$z$s. 
Moreover, the 2D method identifies two additional small substructures to the east and southeast of the center of A4038, around a projected distance of $2R_{200}$, which when using the 3D  phase-space technique, are classified to belong to the central cluster. Given that the 2D technique does not take into account the galaxies within $R_{200}$ (by default) and given the complex nature of this system (with three central groups/clusters in the same line of sight), this is not surprising. The largest substructures outside $5 R_{200}$ (blue and pink) also match when comparing the 2D and 3D techniques.

A few of the detected subgroups within $5 R_{200}$ exhibit significant differences in projected distance and line-of-sight velocity. For example, the substructure located to the east of the center of the cluster (identified consistently by all applied techniques and samples, shown in light green in Figure \ref{calsagos_DSPlus}) is offset by more than $5.7 h^{-1}$ Mpc in projection and displays a velocity difference of $-421$ km s$^{-1}$. These characteristics suggest that it may correspond to a recently accreted group 
currently infalling into the A4038 or A4049 clusters. 
The dark green substructure south of the center of A4038, visible in all three panels of the same figure, shows a velocity offset of $\sim1500$ km s$^{-1}$ and is probably associated with A4038B. Similarly, the smaller northern substructure (pink in the first panel of Figure \ref{calsagos_DSPlus}) presents a velocity difference of $\sim1500$ km s$^{-1}$ and is therefore also likely related to A4038B. These probably correspond to the overdensity seen in the upper panel of Figure \ref{central}, at $|\Delta V|/\sigma = 2.2$ and $R/R_{200} = 1.7$. In future work, a more detailed kinematic analysis of the cluster and its surrounding structures will be presented.

The spectroscopic analysis of the central part of A4038 showed 3 large subgroups: A4038, A4038B and A4049. A4038B lies in the background, spatially coincident with A4038, but with a velocity difference of about 1800~km s$^{-1}$, while A4038 and A4049 have nearly the same central redshifts but are separated by about $2~h^{-1}$ Mpc in projection. Given the complex nature of the central region, a few of the substructures identified in the field of A4038 may in fact belong to one of the other two clusters. More details on the analysis of the dynamics and structure of this cluster, including the determination of filaments and their relation to galaxy properties, will be presented in a future paper.

\subsection{On the color and luminosity distributions of galaxies around A4038}

The ability to probe faint group/cluster populations is a particular strength of deep photometric surveys. In SCALE studies, photometric members are especially valuable, since S-PLUS photo-$z$s have precisions better than 2.5\% for $r < 19.5$ mag, reaching magnitudes significantly fainter than typical spectroscopic surveys and enabling comprehensive galaxy color studies.

Having established in Sections \ref{sec:A4038-3d} to \ref{sec:central} the presence of substructures within A4038, in Section \ref{sec:color-segregation}, we investigate how the dynamical complexity of this system manifests in the color and luminosity distributions of galaxies in and around the cluster. The results of this analysis lead to several important conclusions.

The results shown in the left panel of Figure \ref{cdf} are consistent with the well-established trend in galaxy clusters whereby luminous galaxies are systematically redder than their fainter counterparts, reflecting more efficient quenching in massive systems that host older stellar populations. Lower-luminosity galaxies, in contrast, retain bluer colors due to lower metallicities and/or ongoing or recent star formation.

The radial trends shown in the right panel of Figure~\ref{cdf} suggest that faint galaxies preferentially populate the outskirts of the cluster, while intermediate-luminosity systems are more centrally concentrated. However, the absence of statistically significant differences in the overall radial distributions indicates that luminosity alone does not strongly determine present-day clustercentric location, particularly within the core.

The color–concentration analysis shown in Figure \ref{colorconc} confirms the classical morphology–color relation, in which galaxies with higher concentration indices, typically early-type systems, are redder than less concentrated galaxies. However, the strong magnitude dependence of this relation reveals an important deviation. At low luminosities, galaxies with different structural properties converge toward similar blue colors. We speculate that many low-mass galaxies may sustain some level of star formation regardless of their structural concentration.

Our finding that low-mass galaxies exhibit similar blue colors irrespective of concentration is consistent with results from spatially resolved spectroscopic analyses. Using MaNGA data, \citet{peterken2021size} showed that present-day morphology has a weaker effect on inferred star formation history than color, particularly at low stellar masses. In this regime, morphological differences are therefore not expected to translate into strong differences in recent star formation activity, naturally leading to similar optical colors across structurally distinct systems.

The lack of color separation may also reflect the limited effectiveness of simple structural parameters as morphological proxies at low masses. \citet{lange2015galaxy} demonstrated that the bimodality in the Sérsic index and related concentration measures becomes less pronounced below $M_\star \sim 10^{10}\,M_\odot$, with substantial overlap between early- and late-type systems. Also, \citet{Schombert2018Colors} showed that dwarf ellipticals follow a distinct color–magnitude relation from normal ellipticals, characterized by a systematic blueward deviation, often attributed to younger mean stellar ages in lower-mass systems. In the Virgo cluster, as an example, \citet{urich2017young} has explained the blue early-type dwarfs as ram-pressure stripped galaxies with star formation in their cores.

Finally, recent studies have shown that environmental processes such as ram pressure can induce short bursts of star formation prior to complete gas stripping, producing blue dwarf galaxies that are not currently star-forming but have experienced recent starbursts \citep{grishin2021transforming,grishin2026ram}. Such post-starburst dwarfs have been identified in the Coma and Virgo clusters and can appear blue without ongoing star formation, offering an alternative explanation for the blue colors of low-concentration, low-luminosity galaxies in our sample. A detailed discrimination between ongoing star formation and post-starburst populations would require spectroscopic measurements (e.g., of H$\delta$ absorption), which is beyond the scope of this photometric study.

\section{Conclusions} \label{sec:conclusions}

This paper presents {\bf S}-PLUS {\bf C}lusters {\bf A}nd {\bf L}arge-scale {\bf E}nvironments (SCALE), a new project that focuses on the study of nearby clusters and groups of galaxies out to $5 R_{200}$, taking advantage of the improved photo-$z$s provided by the 12-band photometric system used by S-PLUS. 

S-PLUS DR5 has covered 4,500 deg$^2$ of the southern sky, providing deep, calibrated photometry and photo-$z$s. The spectral information embedded in the 12-band photometry enables the identification of galaxy populations and the measurement of their photo-$z$s with greater precision than that achieved by using broad bands only. This is particularly important for determining the membership of galaxies in clusters and their outskirts \citep{Molino+20}.

In this paper, we revisit a sample of 83 known groups and clusters with {$0.008 \le z \le 0.1$}, selected from the literature with spectroscopic redshifts. Mass, radius, velocity dispersion, and numbers of spectroscopic members are listed in Table\,\ref{tab:59clusters}. Table 2 presents more than 50 measured galaxy parameters, derived either in this work or from the literature, for spectroscopic members of the 83 clusters as well as for interlopers.

Future work will build on this study by systematically identifying galaxy clusters in S-PLUS DR5 through positional matches with eROSITA X-ray sources (Paper II). In this sense, our work establishes a reference sample of well-studied systems, while Paper II extends the approach to a broader, X-ray-selected cluster population (Paper II), providing a complementary perspective on galaxy systems and their large-scale environments.

The usefulness of the photo-$z$s for selecting new members of the clusters and groups was verified in several ways: (1) by presenting several metrics, $\sigma_{\text{NMAD}}$, bias and the distribution of the PDFs, and showing their good performance in Figure\,\ref{fig:photo-z-summary}; (2) by showing the matching distributions of spectroscopic redshifts and photo-$z$s in Figure\,\ref{fig:zoffset}; and (3) by demonstrating the good agreement when substructures are detected using spectroscopic redshifts only or spectroscopic redshifts and photo-$z$s in Figure\,\ref{calsagos_DSPlus}. 

It is worth noticing that the determination of accurate photo-$z$s for the galaxies in the field of the 83 groups and clusters in our sample disclosed a significant number of faint galaxies with no spectroscopy that were at the group/cluster redshift and that can be potential members.
S-PLUS photo-$zs$ then represent an important parameter to guide further spectroscopic surveys of the studied groups and clusters, such as those planned by the 4MOST/CHANCES collaboration \citep{Sifon2025}\footnote{Twelve-band photometry, morphological parameters, and reduced and calibrated images for a subset of the clusters listed in Table \ref{tab:59clusters} were provided to the 4MOST/CHANCES collaboration in 2022-2024 for confection of their spectroscopic target lists, using S-PLUS photo-$zs$. Future spectroscopy to be done with 4MOST will help further refine membership \citep{Mendez-Hernandez2026} specially of the faintest members of these clusters.}.   

The RPM technique was applied to the cluster A4038 to find photometric candidates. These, together with the spectroscopic redshifts available in the literature, were used for the study of the cluster's substructures, comparing techniques that use 2D and 3D phase-space information (with and without photo-$z$s), and obtaining a good agreement in the results.  We find, for the first time, about a dozen substructures within $10 \times R_{200}$ of A4038/A4038B/A4049, extending the results of \citet{Burgett2004}, who studied substructures within $1.3 R_{200}$ of A4038.

The identification of photometric cluster members allowed the segregation analysis presented in Section \ref{sec:color-segregation}. Our results demonstrate that brighter galaxies are systematically redder and more concentrated, while fainter objects are bluer and show extended distributions, as expected. 

The inclusion of a new parameter in the analysis, the galaxy's central concentration, reveals a more nuanced picture: the color–magnitude relation exhibits markedly different slopes for concentrated ($\Delta(g-r) = 0.30$ mag) and extended ($\Delta(g-r) = 0.17$ mag) galaxies, with the largest color differences ($\Delta(g-r) = 0.15$ mag) emerging only for the most luminous systems. Thus, while the well-established morphology–color relation holds for massive galaxies, it progressively weakens at low luminosities, where both concentrated and diffuse systems display significant morphological diversity without a strong color segregation, suggesting similar evolutionary pathways. This convergence challenges a simple morphology–color dichotomy and points toward quenching mechanisms whose efficiency depends on galaxy mass. 
However, as this study is currently limited to a single cluster, the statistical uncertainties and potential environmental biases remain significant.

Overall, these findings indicate that galaxy evolution in cluster environments is regulated by mass-dependent processes, with a clear morphology–color separation emerging only above a critical luminosity threshold. A future study incorporating data from a larger sample of clusters will improve the statistics and enable a robust determination of the stellar mass scale at which color and structural parameters decouple, as well as a statistical characterization of the dominant morphological populations across this transition as a function of environment.

These findings establish the foundation for comprehensive studies of all SCALE clusters, providing robust constraints on environmental processing in diverse cluster environments.

\section*{Data Availability}
The catalogs presented in this work, Tables \ref{tab:59clusters} and \ref{tab:column_description}, are publicly available in the online version of this paper. These tables are also available in multiple file formats through the project website at \url{https://splus-scale.github.io}.

\begin{appendix}

\section{Membership catalog} \label{sec:sup-tables}

The membership catalog comprises all spectroscopically confirmed cluster members and interlopers associated with the clusters listed in Table~\ref{tab:59clusters}, identified using the shifting-gapper technique described in Section~\ref{sec:Sample}. A detailed description of the catalog columns is provided in Table~\ref{tab:column_description}. The complete catalog is available in the online version of this paper. To facilitate accessibility and reuse, we also provide the catalog in multiple file formats through the project website (\url{https://splus-scale.github.io}).

\startlongtable
\begin{deluxetable*}{rllp{0.55\linewidth}}\label{tab:column_description}
\tablecaption{%
Description of each column of membership catalog. The columns are listed in the order of appearance in the table.
}
\tablehead{%
\colhead{No.} & \colhead{Unit} & \colhead{Column name} & \colhead{Column description}
}
\startdata
1 & ~ & cluster & Cluster or group identifier \\
2 & deg & ra & Right Ascension (J2000) in decimal degrees \\
3 & deg & dec & Declination (J2000) in decimal degrees \\
4 & ~ & flag\_member & Membership flag (0=member, 1=interloper) \\
5 & km/s & velocity & Velocity derived from the spectroscopic redshift \\
6 & km/s & velocity\_err & Velocity error \\
7 & km/s & velocity\_offset & Velocity difference relative to the cluster central velocity \\
8 & deg & dist\_proj & Sky projected angular distance between the object and the corresponding cluster center \\
9 & Mpc & dist & Linear distance between the object and the corresponding cluster center \\
10 & ~ & dist\_r200 & Linear clustercentric distance to the object normalized by the corresponding cluster's virial radius (R$_{200}$). \\
11 & ~ & sp\_prob\_gal & Probability of the object being a galaxy \citep{{Nakazono+21}}. \\
12 & deg & sp\_A & Semi-major axis of the object's light distribution, corresponding to its maximum spatial dispersion. \\
13 & deg & sp\_B & Semi-minor axis of the object's light distribution, corresponding to its minimum spatial dispersion. \\
14 & deg & sp\_PA & S-PLUS DR5 position angle (counter-clockwise / World-x) \\
15 & ~ & sp\_ellipticity & Ellipticity (A\_IMAGE / B\_IMAGE) \\
16 & deg & sp\_radius\_petro & S-PLUS DR5 petrosian aperture radius \\
17 & deg & sp\_radius\_20 & Radius enclosing 20\% of the total flux \\
18 & deg & sp\_radius\_50 & Radius enclosing 50\% of the total flux \\
19 & deg & sp\_radius\_90 & Radius enclosing 90\% of the total flux \\
20 & mag arcsec$^{-2}$ & sp\_mu\_max\_g & Instrumental peak surface brightness above background in g \\
21 & mag arcsec$^{-2}$ & sp\_mu\_max\_r & Instrumental peak surface brightness above background in r \\
22 & ~ & sp\_background\_g & Instrumental background at centroid position in g \\
23 & ~ & sp\_background\_r & Instrumental background at centroid position in r \\
24 & ~ & sp\_s2n\_g\_auto & Signal-to-noise ratio of g (auto) \\
25 & ~ & sp\_s2n\_r\_auto & Signal-to-noise ratio of r (auto) \\
26 & mag & sp\_mag\_u\_auto & S-PLUS DR5 magnitude in u band with auto aperture \\
27 & mag & sp\_mag\_g\_auto & S-PLUS DR5 magnitude in g band with auto aperture \\
28 & mag & sp\_mag\_r\_auto & S-PLUS DR5 magnitude in r band with auto aperture \\
29 & mag & sp\_mag\_i\_auto & S-PLUS DR5 magnitude in i band with auto aperture \\
30 & mag & sp\_mag\_z\_auto & S-PLUS DR5 magnitude in z band with auto aperture \\
31 & mag & sp\_mag\_F378\_auto & S-PLUS DR5 magnitude in F378 band (Balmer jump / [O\textsc{ii}]) with auto aperture \\
32 & mag & sp\_mag\_F395\_auto & S-PLUS DR5 magnitude in F395 band (Ca H + K) with auto aperture \\
33 & mag & sp\_mag\_F410\_auto & S-PLUS DR5 magnitude in F410 band (H$\delta$) with auto aperture \\
34 & mag & sp\_mag\_F430\_auto & S-PLUS DR5 magnitude in F430 band (G band) with auto aperture \\
35 & mag & sp\_mag\_F515\_auto & S-PLUS DR5 magnitude in F515 band (Mg b triplet) with auto aperture \\
36 & mag & sp\_mag\_F660\_auto & S-PLUS DR5 magnitude in F660 band (H$\alpha$) with auto aperture \\
37 & mag & sp\_mag\_F861\_auto & S-PLUS DR5 magnitude in F861 band (Ca triplet) with auto aperture \\
38--49 & mag & sp\_mag\_\texttt{[band]}\_PStotal & S-PLUS DR5 total magnitude in \texttt{[band]} band, where a correction to the 3'' aperture has been made taking into account stellar profiles; \texttt{[band]} is one of: u, g, r, i, z, F378, F395, F410, F430, F515, F660, F861 \\
50--61 & mag & sp\_mag\_\texttt{[band]}\_aper\_6 & S-PLUS DR5 magnitude in \texttt{[band]} band within 6'' aperture; \texttt{[band]} is one of: u, g, r, i, z, F378, F395, F410, F430, F515, F660, F861 \\
62--73 & mag & sp\_mag\_err\_\texttt{[band]}\_auto & S-PLUS DR5 magnitude error in \texttt{[band]} band; \texttt{[band]} is one of: u, g, r, i, z, F378, F395, F410, F430, F515, F660, F861 \\
74--85 & mag & sp\_mag\_err\_\texttt{[band]}\_PStotal & S-PLUS DR5 magnitude error in \texttt{[band]} band for PStotal magnitude; \texttt{[band]} is one of: u, g, r, i, z, F378, F395, F410, F430, F515, F660, F861 \\
86--97 & mag & sp\_mag\_err\_\texttt{[band]}\_aper\_6 & S-PLUS DR5 magnitude error in \texttt{[band]} for 6''-aperture magnitude; \texttt{[band]} is one of: u, g, r, i, z, F378, F395, F410, F430, F515, F660, F861 \\
98--101 & mag & sp\_mag\_g\_\texttt{[type]} & S-PLUS DR5 magnitude in g band for \texttt{[type]} magnitude; \texttt{[type]} is one of: aper\_3, res, iso, petro \\
102--105 & mag & sp\_mag\_r\_\texttt{[type]} & S-PLUS DR5 magnitude in r band for \texttt{[type]} magnitude; \texttt{[type]} is one of: aper\_3, res, iso, petro \\
106 & ~ & sp\_field & Survey field identifier \\
107 & ~ & sp\_photoz & S-PLUS DR5 single-point photometric redshift estimate \citep{{Lima+22}} \\
108 & ~ & sp\_photoz\_odds & Area of PDF within 0.02 of the PDF peak (photo-z odds) \\
109--111 & ~ & sp\_photoz\_pdf\_weights\_\texttt{[i]} & Weights of the gaussian components of the photometric-redshift PDF mixture, where \texttt{[i]} is one of 0, 1, 2 \\
112--114 & ~ & sp\_photoz\_pdf\_means\_\texttt{[i]} & Means of the gaussian components of the photometric-redshift PDF mixture, where \texttt{[i]} is one of 0, 1, 2 \\
115--117 & ~ & sp\_photoz\_pdf\_stds\_\texttt{[i]} & Standard deviations of the gaussian components of the photometric-redshift PDF mixture, where \texttt{[i]} is one of 0, 1, 2 \\
118 & ~ & sp\_in\_overlap\_region & Flag indicating object lies in overlap region (bool/int) \\
119--126 & mag & ls\_mag\_\texttt{[band]} & Legacy Survey DR10 magnitudes where \texttt{[band]} is one of: g, r, i, z, w1, w2, w3, w4 \\
127 & ~ & ls\_type & Legacy Survey DR10 morphological type \\
128 & arcsec & ls\_shape\_r & Legacy Survey effective radius (arcsec) \\
129 & ~ & lit\_redshift & Spectroscopic redshift \citep{{Lima+24}} \\
130 & ~ & lit\_redshift\_err & Spectroscopic redshift error \\
131 & ~ & lit\_class\_spec & Spectroscopic classification \\
132 & ~ & lit\_original\_class\_spec & Original spectroscopic classification before grouping \\
133 & ~ & lit\_source & Catalogue/source of the spectroscopic redshift \\
134 & ~ & lit\_redshift\_flag & Flag indicating spectroscopic redshift quality \\
\enddata
\tablecomments{Columns prefixed with \texttt{sp\_} (10--117) are drawn from S-PLUS DR5 (E. V. Lima et al. 2026, submitted), while those prefixed with \texttt{ls\_} (118--127) originate from Legacy Survey DR10 \citep{legacy}. Columns prefixed with \texttt{lit\_} (128---133) originate from redshift compilation from the literature \citep{Lima+24}. Columns without these prefixes (1--9) were derived in this work. All magnitudes are reported in the AB system. This table is published in its entirety in the electronic edition of the {\it Astrophysical Journal} and is also available in several digital formats in the project's website (\url{https://splus-scale.github.io}). The columns description is shown here for guidance regarding its form and content.}
\end{deluxetable*}

\section{Mass comparison}
\label{sec:mass-comparison}
\begin{figure}[!ht]
    \centering
    \includegraphics[width=0.5\linewidth]{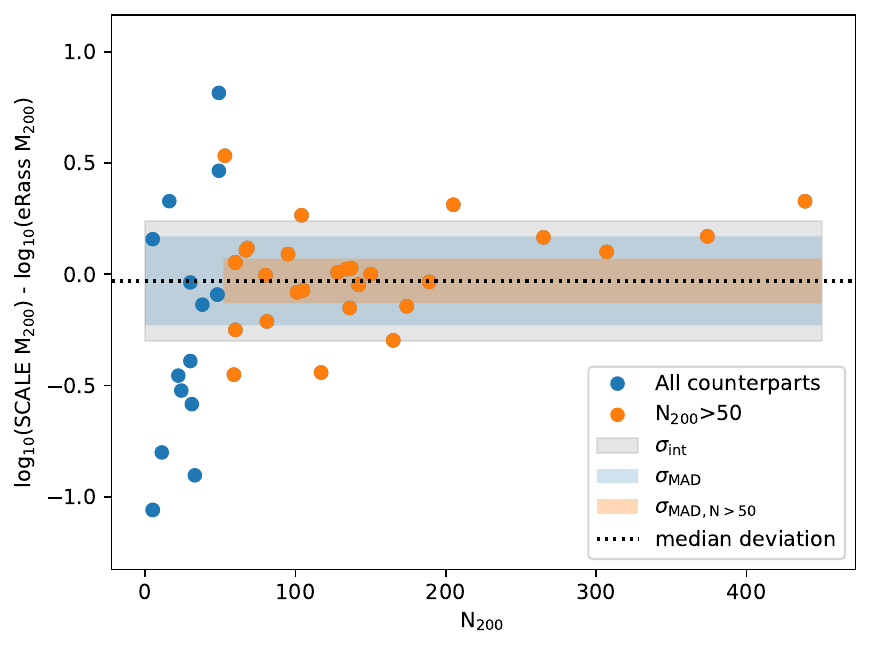}
    \caption{Logarithmic mass ratio between dynamical mass estimates from the SCALE sample and eROSITA re-scaled masses (i.e., eROSITA $M_{500}$  were converted to $M_{200}$, to allow a comparison - see text for details). The ratio shows a small median offset of $\sim0.03$ dex, and $\sigma_{\rm MAD} = 0.20$ dex. Better agreement is observed if considered only structures with $N_{200}>50$. }
    \label{fig:mass_comp}
\end{figure}

Figure\,\ref{fig:mass_comp} presents the comparison between the dynamical mass estimates derived in this paper, as discussed in Section\,\ref{sec:Sample}, with masses derived from eROSITA X-ray observations \citep{Bulbul2024}. We convert $M_{500}$ given in the eRosita catalog to $M_{200}$ assuming a Navarro-Frenk-White halo profile \citep{Navarro1996} with a fixed concentration $c=4.5$ \citep{Duffy2008}. The $49$ matched systems were obtained using the following conditions: with $\Delta R<0.5~h^{-1}$ Mpc and $\Delta z < 0.005$.

As previously discussed, the small median offset and moderate intrinsic scatter indicate that the two mass estimates are statistically consistent within their expected uncertainties. We note a larger offset of $\sim0.2$ dex for rich clusters, which might be biased due to the conversion between masses ($M_{500} \to M_{200}$).

Further comparisons with other dynamical mass studies are challenging, as relatively few systems in the literature have adequate spectroscopic sampling in the southern hemisphere. As a result, even when cross-matching with existing catalogs, the overlap with our sample remains limited.

\end{appendix}

\section*{Acknowledgments}
The S-PLUS project, including the T80-South robotic telescope and the S-PLUS scientific survey, was founded as a partnership between the Fundação de Amparo à Pesquisa do Estado de São Paulo (FAPESP), the Observatório Nacional (ON), the Federal University of Sergipe (UFS), and the Federal University of Santa Catarina (UFSC), with important financial and practical contributions from other collaborating institutes in Brazil, Chile (Universidad de La Serena), and Spain (Centro de Estudios de Física del Cosmos de Aragón, CEFCA). We further acknowledge financial support from the São Paulo Research Foundation (FAPESP), the Brazilian National Research Council (CNPq), the Coordination for the Improvement of Higher Education Personnel (CAPES), the Carlos Chagas Filho Rio de Janeiro State Research Foundation (FAPERJ), and the Brazilian Innovation Agency (FINEP).

The S-PLUS collaboration members are grateful for the contributions from CTIO staff in helping in the construction, commissioning, and maintenance of the T80-South telescope and camera. We are also indebted to Rene Laporte, INPE, and Keith Taylor for their essential contributions to the project. From CEFCA, we particularly would like to thank Antonio Marín-Franch for his invaluable contributions in the early phases of the project, David Cristóbal-Hornillos and his team for their help with the installation of the data reduction package JYPE version 0.9.9, César Íñiguez for providing 2D measurements of the filter transmissions, and all other staff members for their support with various aspects of the project.

The Legacy Surveys consist of three individual and complementary projects: the Dark Energy Camera Legacy Survey (DECaLS; Proposal ID \#2014B-0404; PIs: David Schlegel and Arjun Dey), the Beijing-Arizona Sky Survey (BASS; NOAO Prop. ID \#2015A-0801; PIs: Zhou Xu and Xiaohui Fan), and the Mayall z-band Legacy Survey (MzLS; Prop. ID \#2016A-0453; PI: Arjun Dey). DECaLS, BASS and MzLS together include data obtained, respectively, at the Blanco telescope, Cerro Tololo Inter-American Observatory, NSF’s NOIRLab; the Bok telescope, Steward Observatory, University of Arizona; and the Mayall telescope, Kitt Peak National Observatory, NOIRLab. Pipeline processing and analyses of the data were supported by NOIRLab and the Lawrence Berkeley National Laboratory (LBNL). The Legacy Surveys project is honored to be permitted to conduct astronomical research on Iolkam Du'ag (Kitt Peak), a mountain with particular significance to the Tohono O’odham Nation.

DESI construction and operations is managed by the Lawrence Berkeley National Laboratory. This research is supported by the U.S. Department of Energy, Office of Science, Office of High-Energy Physics, under Contract No. DE–AC02–05CH11231, and by the National Energy Research Scientific Computing Center, a DOE Office of Science User Facility under the same contract. Additional support for DESI is provided by the U.S. National Science Foundation, Division of Astronomical Sciences under Contract No. AST-0950945 to the NSF’s National Optical-Infrared Astronomy Research Laboratory; the Science and Technology Facilities Council of the United Kingdom; the Gordon and Betty Moore Foundation; the Heising-Simons Foundation; the French Alternative Energies and Atomic Energy Commission (CEA); the National Council of Science and Technology of Mexico (CONACYT); the Ministry of Science and Innovation of Spain, and by the DESI Member Institutions. The DESI collaboration is honored to be permitted to conduct astronomical research on Iolkam Du’ag (Kitt Peak), a mountain with particular significance to the Tohono O’odham Nation.

C.M.d.O. thanks the Fundação de Amparo à Pesquisa do Estado de São Paulo (FAPESP) for funding through the grants 2019/26492-3 and Conselho Nacional de Desenvolvimento Científico e Tecnológico (CNPq) through grant 309209/2019-6. N.M.C. thanks the Coordenação de Aperfeiçoamento de Pessoal de Nível Superior -- Brasil (CAPES) -- Finance Code 88887.133104/2025-00. P.A.A.L. thanks the support from CNPq, grants 310260/2025-6 and 404160/2025-5, and FAPERJ, grant E-26/200.545/2023. A.L.B.R. thanks the support from CNPq, grants 316317/2021-7 and 404160/2025-5. L.S.J. acknowledges the support from CNPq (308994/2021-3) and FAPESP (2011/51680-6). D.E.O-R. acknowledges financial support from ANID Fondo ALMA 2025 project 31250033. D.E.O-R. also thanks ANID InESGénero project INGE210025.
E.V.R.L. acknowledges the financial support given by CAPES (grant 88887.470064/2019-00), CNPq (grant 169181/20170), and FAPESP (grant 2024/15229-8). L.D acknowledges FAPESP grants 2024/03575-9 and 2025/11378-1.   A.V.S.C acknowledge financial support from the Consejo Nacional de Investigaciones Científicas y Técnicas (CONICET), Agencia I+D+i (PICT 2019–03299) and Universidad Nacional de La Plata (Argentina). A.V.S.C. also thanks FAPESP for the support through Grant 2025/05085-1. R.C.F. acknowledges support from CNPq (Grant No. 302270/2018-3). R.D. gratefully acknowledges support by the ANID BASAL project FB210003. L.N. acknowledges FAPESP grant number 2024/07281-0. A.P.C. thanks the financial support from CAPES - Finance Code 001. C.L. acknowledges support from FCT, Fundação para a Ciência e a Tecnologia, through national funds, grant UIDB/04434/2025. G.M. gratefully acknowledges FAPESP for the support grant 2024/10923-3 and 2025/14602-0.  C.L.D. acknowledges support from the Agencia Nacional de Investigación y Desarrollo (ANID) through Fondecyt project 3250511. The work of E.R.C. is supported by the international Gemini Observatory, a program of NSF NOIRLab, which is managed by the Association of Universities for Research in Astronomy (AURA) under a cooperative agreement with the U.S. National Science Foundation, on behalf of the Gemini partnership of Argentina, Brazil, Canada, Chile, the Republic of Korea, and the United States of America. M.S.C. acknowledges funding from FAPESP grant 2025/12629-8. A.R.L. acknowledges financial support from FAPESP through grant 2025/09544-0, and the support from Consejo Nacional de Investigaciones Científicas y Técnicas (CONICET). F.A.-F. acknowledges support from FAPESP grant 2024/00822-5.

\bibliographystyle{aasjournal}
\bibliography{lib.bib}

\end{document}